%% file: at3dmeg1.rev.v1b.tex
\documentclass[useAMS,usenatbib]{mn2e}
\include{abbrev}

\usepackage{graphics,epsf,epsfig,graphicx}
\usepackage[total={17.8cm,24.0cm},centering]{geometry}
\usepackage{times}

\title[The merger origin of NGC~680 and NGC~5557]{The \AD\ project -- IX.  The merger origin of  a fast and a slow rotating Early-Type Galaxy revealed with deep optical imaging: first results}

\author
[
Pierre-Alain Duc et al.]{\parbox{\textwidth}{
Pierre-Alain Duc,$^{1}$\thanks{E-mail:\texttt{paduc@cea.fr}}
Jean-Charles Cuillandre,$^{2}$
Paolo Serra,$^{3}$
Leo Michel-Dansac,$^{4}$
Etienne Ferriere,$^{1}$
Katherine Alatalo,$^{5}$
Leo Blitz,$^{5}$
Maxime Bois,$^{6,4}$
Fr\'ed\'eric Bournaud,$^{1}$
Martin Bureau,$^{7}$
Michele Cappellari,$^{7}$
Roger L. Davies,$^{7}$
Timothy A. Davis,$^{7}$
P. T. de Zeeuw,$^{6,8}$
Eric Emsellem,$^{6,4}$
Sadegh Khochfar,$^{9}$
Davor Krajnovi\'c,$^{6}$
Harald Kuntschner,$^{10}$
Pierre-Yves Lablanche,$^{4}$
Richard M. McDermid,$^{11}$
Raffaella Morganti,$^{3,12}$
Thorsten Naab,$^{13}$
Tom Oosterloo,$^{3,12}$
Marc Sarzi,$^{14}$
Nicholas Scott,$^{7}$
Anne-Marie Weijmans,$^{15}$\thanks{Dunlap Fellow}
and Lisa M. Young $^{16}$}\vspace{0.4cm}\\ 
\parbox{\textwidth}{$^{1}$Laboratoire AIM Paris-Saclay, CEA/IRFU/SAp, CNRS/INSU, Universit\'e Paris Diderot, 91191 Gif-sur-Yvette Cedex, France\\
$^{2}$Canada-France-Hawaii Telescope Corporation,	65-1238 Mamalahoa Hwy., Kamuela, Hawaii 96743 USA    \\
$^{3}$Netherlands Institute for Radio Astronomy (ASTRON), Postbus 2, 7990 AA Dwingeloo, The Netherlands\\
$^{4}$ Centre de Recherche Astrophysique de Lyon, Universit\'e Lyon 1, Observatoire de Lyon,  Ecole Normale Sup\'erieure de Lyon, CNRS, UMR 5574, 9 avenue Charles Andr\'e, F-69230 Saint-Genis Laval, France\\
$^{5}$Department of Astronomy, Campbell Hall, University of California, Berkeley, CA 94720, USA\\
$^{6}$European Southern Observatory, Karl-Schwarzschild-Str. 2, 85748 Garching, Germany\\
$^{7}$Sub-department of Astrophysics, Department of Physics, University of Oxford, Denys Wilkinson Building, Keble Road, Oxford OX1 3RH\\
$^{8}$Sterrewacht Leiden, Leiden University, Postbus 9513, 2300 RA Leiden, the Netherlands\\
$^{9}$Max Planck Institut f\"ur extraterrestrische Physik, PO Box 1312, D-85478 Garching, Germany\\
$^{10}$Space Telescope European Coordinating Facility, European Southern Observatory, Karl-Schwarzschild-Str. 2, 85748 Garching, Germany\\
$^{11}$Gemini Observatory, Northern Operations Centre, 670 N. A`ohoku Place, Hilo, HI 96720, USA\\
$^{12}$Kapteyn Astronomical Institute, University of Groningen, Postbus 800, 9700 AV Groningen, The Netherlands\\
$^{13}$Max-Planck-Institut f\"ur Astrophysik, Karl-Schwarzschild-Str. 1, 85741 Garching, Germany\\
$^{14}$Centre for Astrophysics Research, University of Hertfordshire, Hatfield, Herts AL1 9AB\\
$^{15}$Dunlap Institute for Astronomy \& Astrophysics, University of Toronto, 50 St. George Street, Toronto, ON M5S 3H4, Canada\\
$^{16}$Physics Department, New Mexico Institute of Mining and Technology, Socorro, NM 87801, USA
}}

\begin{document}

\date{Accepted: 26/05/2010}


\maketitle
\clearpage
\newpage

\label{firstpage}

\begin{abstract}
The mass assembly of galaxies leaves imprints in their outskirts, such as shells and tidal tails. The frequency and properties of such fine structures depend on the main acting mechanisms -- secular evolution, minor or major mergers -- and on the age of the last substantial accretion event. We use this to constrain the mass assembly history of two apparently relaxed nearby Early-Type Galaxies (ETGs) selected from the \AD\ sample, NGC~680 and NGC~5557. Our ultra deep optical images obtained with MegaCam on the Canada-France-Hawaii Telescope  reach 29~\sbr in the g--band. They  reveal very low-surface brightness (LSB) filamentary structures around these ellipticals. Among them, a gigantic 160~kpc long, narrow, tail East of NGC~5557 hosts three gas-rich star-forming objects, previously detected in HI with the Westerbork Synthesis Radio Telescope and in UV with GALEX. NGC~680 exhibits two major diffuse plumes apparently connected to extended HI tails, as well as a series of arcs and shells. Comparing the outer stellar and gaseous morphology of the two ellipticals with that predicted from models of colliding galaxies, we argue that the LSB features are tidal debris and that each of these two ETGs was assembled during a relatively recent, major wet merger, which most likely occurred after the redshift $z \simeq 0.5$ epoch.
Had these mergers been older, the tidal features should have already fallen back or be destroyed by more recent accretion events.
However the absence of molecular gas and of a prominent young stellar population in the core region of the galaxies indicates that the merger is at least 1-2 Gyr old: the memory of any merger-triggered nuclear starburst has indeed been lost. The star-forming objects found towards the collisional debris of NGC~5557 are then likely Tidal Dwarf Galaxies. Such recycled galaxies here appear to be long-lived and continue to form stars while any star formation activity has stopped in their parent galaxy. The inner kinematics of NGC~680 is typical for fast rotators which make the bulk of nearby ETGs in the \AD\ sample. 
On the other hand, NGC~5557 belongs to the poorly populated class of massive, round, slow rotators that are predicted by  semi-analytic models and cosmological simulations to be the end-product of a complex mass accretion history, involving ancient major mergers and more recent minor mergers. Our observations suggest that under specific circumstances a  single binary merger  may dominate the formation history of such objects and thus that at least  some massive ETGs may form at relatively low redshift. Whether the two galaxies studied here are representative of their own sub-class of ETGs is still an open question that will be addressed by an on-going deep optical survey of  \AD\   galaxies.
 \end{abstract}

\begin{keywords}
galaxies: formation --
galaxies: elliptical and lenticular --
galaxies: interactions --
galaxies: individual (NGC~680) --
galaxies: individual (NGC~5557)
\end{keywords}

\section{Introduction}

\label{sect:origin}
According to standard hierarchical cosmological models,  the formation of galaxies is a gradual process based on the assembly of building blocks making their mass, and on the  accretion of gas making their stars. The details of this scenario are however still under active discussion. The debate has been specifically hot for the population of Early-Type Galaxies (ETGs), since, being the most massive systems,  they are believed to be the end product of the mass assembly process. 
Since the seminal simulation by \cite{Toomre77a} of a spiral-spiral collision and follow-up works \citep[e.g.][]{Negroponte83} showing that the merger remnant had  a de Vaucouleurs profile \citep[e.g.][]{Hibbard99b}, much attention has been given to the merger scenario for the formation of early-type galaxies. 
However, the luminosity profile is by far not the only parameter that characterizes galaxies. 
The {\tt SAURON} \citep{deZeeuw02} and \AD\  projects  \citep[][Paper I]{Cappellari11} have recently  provided a wealth of observational data on nearby ETGs of particular use for constraining their nature. In particular, the study of the inner stellar kinematics enabled by integral field spectroscopy put the attention on some little-known properties of ETGs:  the large majority (80--85\%) of  ETGs rotate; a fraction of ETGs, especially among the slow rotating systems, show intriguing features, such as Kinematically Distinct Components (KDCs), including counter rotating cores \citep[][Paper II]{Krajnovic11}. This suggests that although early-type galaxies  globally follow common scaling relations,  they might in fact constitute different families of objects, with different histories for their mass assembly. Complementary to the observational efforts,  various types of numerical simulations have been carried out to reproduce the photometric and kinematical properties of ETGs:   binary major and minor mergers   \citep[among the most recent ones, ][]{Jesseit09,Johansson09,DiMatteo09,Hoffman10,Bois10a},  series of major mergers \citep{Weil96}, minor mergers \citep[e.g.][]{Bournaud07b},  multiple collisions following hierarchical structure formation scenarios \citep[e.g.][]{Naab06,Burkert08} or cosmological simulations with zoom-in techniques  \citep[e.g.][]{Martig09}.
Most of these models successfully form the fast rotators that dominate the bulk of the ETG population. The formation of the slow rotators with single binary mergers turns out to be more difficult \citep[][Paper VI]{Bois11}. Producing the massive, round,  ones  typically located in groups and clusters seems to require a complex mass assembly history that was initiated at high redshift by a major merger \citep[][Paper VIII]{Khochfar11}. \\

Characterizing and dating the mergers that contributed to the mass assembly of galaxies appear thus as fundamental. Among the most valuable and convenient tracers   of mergers are the fine structures around galaxies:   shells,  rings, plumes and tails. Indeed, depending on whether massive  galaxies were formed through a monolithic collapse,  rapid cold-gas accretion followed by violent disk instabilities, minor/major, gas--rich/gas-poor mergers,  the ubiquity of  collisional debris  will go from none to significant. 
Furthermore, relics  like tidal tails are transient structures with surviving ages varying between a few hundred Myr to a few Gyr, depending on their nature and location \citep{Hibbard95b}. Their observations thus provide a clock that can be calibrated using numerical simulations. 

Collisional relics may be traced mapping their hydrogen gas with radio interferometers  or imaging their stellar populations with  optical cameras.
\cite{Malin80} have used optical photographs of ellipticals and  unsharp masking techniques to disclose concentric stellar shells around some massive ellipticals. These fine structures are generally interpreted as imprints of accretion and disruption of low-mass satellites. \cite{Schweizer90} and \cite{Schweizer92b} performed a more systematic survey of fine structures around ETGs, introducing a fine structure index $\Sigma$ which basically characterizes the number of observed distinct features, such as ripples and jets. Unfortunately, collisional debris are generally faint, and those tracing most efficiently major mergers -- the 50-100 kpc long tidal tails -- have a surface brightness that is typically 24--26 \sbr when young and below 27~\sbr when getting older and fading. 
Their detection requires specific observing conditions and techniques: for galaxies  that are resolved in stars, star counts allow to reach  surface brightness limits much below  30~\sbr. At distances greater than a few Mpc, only the integrated light is available with the current facilities. The limiting factors are then not only the exposure time but as importantly the sky background level, the flat  field subtraction, the scattered light from bright stars  and the field of view.  Most efforts in the detection of extremely low surface brightness  structures around galaxies have so far been done for spirals; they have for instance lead  to the discovery of a stellar bridge  linking  the Local Group  galaxies M31 and M33 \citep{McConnachie09}, or spectacular streams around nearby spirals \citep{Martinez-Delgado10}. 

In the hierarchical scenario, ETGs should exhibit even more fine structures than late type galaxies  \citep[see simulations by][]{Johnston08,Peirani10,Michel-Dansac10}. 
Unfortunately  in clusters of galaxies  where ETGs prevail, the fine structures are particularly fragile and are easily destroyed. Multiple stellar streams were nonetheless recently disclosed  by two dedicated deep surveys of the Virgo Cluster with the  Burrell Schmidt 
telescope \citep{Mihos05,Janowiecki10} and the Canada-France-Hawaii-Telescope (CFHT) as part of the Next Generation Virgo Cluster Survey (NGVS; Ferrarese et al., 2011, in prep.). \cite{Tal09} have however confirmed that streams are more frequent in field ETGs where they should be longer lived.  
\cite{vanDokkum05} collected on deep optical images of cosmological fields a large sample of red anonymous ETGs located at a median redshift of 0.1 and identified around them tidal debris. He claimed that  70\% of  the bulge-dominated galaxies show tidal perturbations telling about a recent merger.
In parallel, the frequent presence of gaseous streams  around ETGs has been disclosed  by  HI surveys carried out with the Very Large Array \citep[e.g.][]{Schiminovich95,Sansom00}, with the Australia Telescope Compact Array   \citep{Oosterloo07}  or the Westerbork Synthesis Radio Telescope (WSRT) \citep{Oosterloo10}.

Prompted by these theoretical studies and  early observations and capitalizing on the wealth of ancillary data obtained by the \AD\ project, we have initiated a deep imaging survey of  very nearby, well known,  ETGs in isolation or in groups. The observations, carried out with the wide field of view  MegaCam camera installed on the CFHT,  allow to study the morphology of the selected galaxies out to very large radius, obtain maps of their outskirts at unprecedented depth, and study their even larger scale environment. 

We report in this paper results obtained on two ellipticals of the \AD\  sample, NGC~680 and NGC~5557 that belong to the  two principle sub-classes of ETGs: slow and fast rotators.  Both turn out to exhibit  in their surrounding spectacular  tidal features.  The optical observations and data reduction, as well as the data gathering of ancillary  data, including HI maps with the Westerbork Synthesis Radio Telescope (WSRT), are detailed in Sect.~\ref{sec:data}.  The detection  and characterization of their outer fine structures, including Tidal Dwarf Galaxies candidates,  are presented in Sect.~\ref{sec:results} and a link is made  to the inner properties of the two galaxies. Finally in Sect.~\ref{sec:discussion}, we discuss the merger scenario for the origin of these two specific systems, and  address their representativeness among the  \AD\ galaxies. 

\begin{figure*}
\centerline{
 \includegraphics[width=\textwidth]{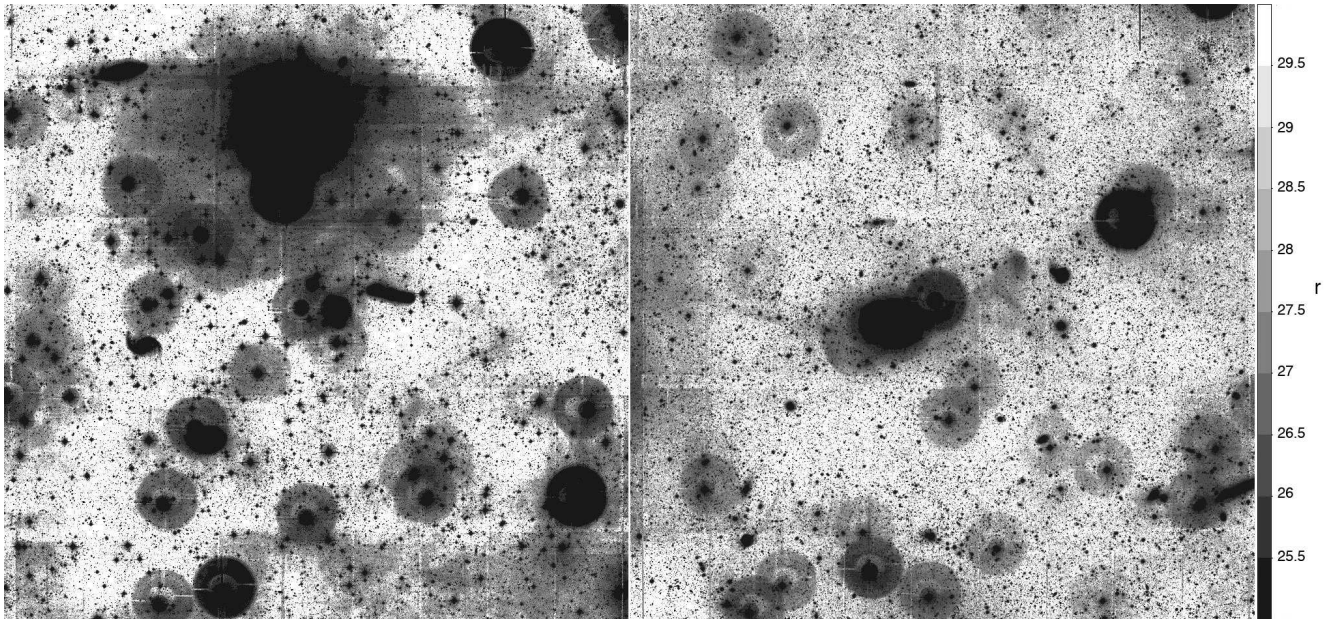}
}
\caption{r--band surface brightness map of the 1 square degree MegaCam fields around NGC~680 (left) and NGC~5557 (right). The two ETGs are located close to the  center of the images. The surface brightness scale is shown to the right. The images are displayed with cuts between 25 and 30~\sbr\ to enhance the structures with the faintest surface brightness: instrument signatures, foreground stars, stellar halos, galactic halos and collisional debris.    At the distance of  NGC~680 and NGC~5557, 1 degree corresponds to  resp.~650 kpc and 670~kpc. North is up and East left.}
\label{fig:deep-field}
\end{figure*}

\section{Observations and data reduction}
\label{sec:data}

\subsection{The two targets}
The two selected targets  presented in this paper, NGC~680 and NGC~5557, lie at roughly the same distance: resp. 37.5 and 38.8 Mpc (Paper I) 
\footnote{There is an uncertainty in the real distance of NGC~5557. The adopted value is the mean of the redshift-independent distance determinations in the NASA/IPAC Extragalactic Database. The latest value determined by \cite{Cantiello05} based on the measure of the Surface Brightness  Fluctuations of the galaxy is 47.9 Mpc. If this distance is true, the sizes of the tidal features detected by MegaCam should be multiplied by a factor of 1.2.}.
Both are classified as very early-type galaxies: $T<-4$ in the Leda classification \citep{Paturel03}, resp. E2 and E1 with the de Vaucouleurs classification.  They are roundish structures, with ellipticities $\epsilon$ between 0.15 and 0.2. 
With \Mk\ of resp. -24.1 and -24.8 \citep{Skrutskie06}, they are massive systems. Having specific angular momenta  \LamR\ of resp.   0.46 and  0.05,  NGC~680 and NGC~5557 unambiguously belong to the classes of resp. fast and slow rotators \citep[][Paper III]{Emsellem11}.
The two galaxies were initially selected as targets for MegaCam observations because of the intriguing presence  of HI  clouds in their vicinity that showed no optical counterparts on the available images at that time (from the Sloan Digital Sky Survey and Isaac Newton Telescope; see details in Paper I). The HI data were obtained with a dedicated survey with the WSRT  made as part of the \AD\ project \citep{Serra11}.  One motivation of the proposed deep optical imaging was  the investigation of the nature of these clouds which could either be primordial accreted clouds or tidal debris. 
A paper based on MegaCam observations using a similar technique as this paper and revealing an optical counterpart to the gigantic HI structure in the Leo ring has been published by \cite{Michel-Dansac10b}.

\subsection{Deep imaging with MegaCam}

\subsubsection{Observations}
Deep optical images of NGC~680 and NGC~5557  were obtained between January and September  2009 with the MegaCam camera installed on
the CFHT. One square degree field of view images were obtained with the g', r' and i' bands. 

Each observation in a given filter consisted of 6 exposures sequenced in time within a 45 minutes time window in order to minimize sky background variations due to the inevitable varying observing environment. Indeed since the true sky varies in the optical by a few percent in shape (gradients) and intensity on timescales of the order  of one hour, exposures used to build a sky must fit within such time  window, and the dithering pattern around the object must be optimized  to integrate the science field at all time.  Since the MegaCam field of view is much larger than the scale of the observed objects,  an extremely large dithering pattern is applied (at least  7' distance from center to center of all 6 images): this allows 100\% of the integration time to be spent on the science field while integrating the background sky all the  while. 
 
Such large dithering patterns ensure that all known optical artifacts plaguing MegaCam, especially the largest stellar halos (6') caused by internal reflections in the  camera optics, can be rejected by building a master sky based on only six exposures. Without such approach, a single frame (or stack made of only slightly dithered frames) exhibits a large radial structure due to the use of twilight flat-fields convolved with a photometric response map offering a flat photometric response across the camera \citep{Magnier04,Regnault09}. 
Such data limits the detection of faint features against the sky background to a surface brigthness of only 27~\sbrs\ over scales of only a few arcminutes.

The number of exposures, total integration time, average seeing and background level are listed in Table~\ref{tab:obsmeg}.
Weather conditions were photometric for all data set.

\begin{table}
\caption{MegaCam observations}
\begin{tabular}{lccccc}
Galaxy & Band & N & Integration & IQ & Backgound \\
              &            &                                 &     s                         & arcsec & ADUs \\
     (1) & (2) & (3) & (4) & (5) & (6) \\         
\hline \\
NGC~680 & g & 6 & 882 & 0.95 & 277 \\
                   & r & 12 & 2760 & 0.71 & 726 \\
                   & i & 6 & 714 & 0.58 & 589 \\
\hline \\
NGC~5557 & g &   6 & 882 & 1.37 & 191 \\
                    &  r  &  6  &   1380 & 1.12 & 364 \\
                    & i  &   6  & 714 & 0.95 & 621 \\        \hline 
\multicolumn{6}{l}{\parbox{\columnwidth}{   (3) Number of individual exposures  (4) Total integration time  (5) Image Quality: FWHM of the PSF (6) Background level}} \\                        
                          
\end{tabular}
\label{tab:obsmeg}
\end{table}

\subsubsection{Data reduction and sensitivity} 
The data were processed at CFHT with the Elixir Low Surface   
Brightness arm of the MegaCam pipeline at CFHT, Elixir-LSB   
(Cuillandre et al. 2011, in prep.). The input data for the Elixir-LSB  
comes from Elixir, the main pipeline which offers a full detrending of  
the individual frames, the correction of the camera photometric  
response, as well as the astrometric  and photometric calibration. The  
output product of Elixir-LSB  are single frames and stacks where the  
sky background has been  modeled and removed, opening the possibility  
for photometry on extremely large scales (arguably the size of the  
field of view) as well as the detection of extremely faint structures  
above the background. Coupled to the proper observing strategy (see  
above), Elixir-LSB achieves in all bands a sky flattening  of 0.2\%,  
nearly 7 magnitudes fainter than the sky background. In the g-band,  
this defines a detection limit at the 29th  magnitude per square  
arcsecond level. While inspired from infrared nodding observing
techniques, the fundamental difference with optical data such as
the MegaCam images is that the sky background is very low and the
image signal is dominated by astronomical sources. Special filtering
and rejection techniques are used in Elixir-LSB in order to reject
these objects and derive the radial shape of the sky background across
the camera field of view induced by the flat-fielding based on twilight
sky exposures convolved with a photometric map correction \citep{Magnier04}.\\

The sky corrected individual images are then finely calibrated for  
astrometry and photometry by the software SCAMP \citep{Bertin06} and  
subsequently resampled, skipping the usual internal sky subtraction,  
by the software SWARP \citep{Bertin02} at  a pixel scale of 0.561  
arcsec per pixel ($3\x3$ binning) in order to boost the signal to  
noise ratio on faint extended objects. The stacks used in the science  
analysis are finally produced by a Elixir-LSB routine optimized for  
low numbers of input frames.\\

The images of the two observed fields, scaled in \sbr, are displayed on Fig.~\ref{fig:deep-field}.
At the depth of the images, many foreground, faint,  stars show up. Given the large extent of the studied structures and their very low surface brightness, the subtraction of the contaminating  light from these stars as well as from distant unresolved galaxies become a critical issue. 
These  sources  were subtracted with a ring--filter technique \citep{Secker95}. 
The  low--surface brightness halos of the brightest  stars, and in general of all luminous objects, including the galaxies,  are prominent on the MegaCam images  and much more difficult to subtract. They are among  the strongest sources of confusion, especially when close to the target galaxies.

\subsubsection{Galaxy subtraction} 
The detection of the inner fine structures required the subtraction of the galaxies. To do this, their light profiles  were modeled in 2D with GALFIT \citep{Peng02}. As a first step, the images were rebinned and the stars and other bright objects -- except the target --  masked. The  masks were created from the images from which the large-scale structures  had been subtracted.  The pixels defining the masks were selected  based on their S/N ratio and a threshold value manually adjusted to ensure the detection of  the faintest stars. The noise and sky levels were then estimated  fitting the histogram of the intensity distribution of the residual images with a Gaussian. 

Both NGC~680 and NGC~5557 were fitted by a de Vaucouleurs profile, i.e. fixing  a Sersic index to 4. Relaxing the value of this index, or adding additional components, did not decrease significantly the residuals. Note that our study mainly addresses collisional debris far out from the ellipticals; thus our results do not suffer much from uncertainties in the galaxy modeling.

\subsection{Ancillary data from \AD\ and GALEX}
Most of the ancillary data used in this study were taken as part of the \AD\ survey and their acquisition and data reduction  described  in dedicated  papers. 
In particular, the HI data shown in this paper were extracted from a survey carried out  with the Westerbork Synthesis Radio Telescope. The FWHM of the WSRT primary beam was 36 arcmin. 
Each galaxy was observed for 12 hours with 1024 frequency channels covering a bandwidth of 20 MHz. This correpsonds to a channel width of $\sim4$ \kms\ over a velocity range of $\sim4000$ \kms. Data were reduced using a pipeline based on the Miriad package \citep{Sault95}. The total-HI images shown in this paper are obtained from data cubes with a FWHM of 51\x 34 arcsec for NGC~680 and  39\x 36 arcsec for NGC~ 5557. The limiting column density is 2.2 and $2.8 \x 10^{19}~\cmm$ for NGC~680 and NGC~5557, respectively. More details on the HI  observations are given in \cite{Serra11}.

 Furthermore deep UV data of NGC~5557 obtained in July 2007 were queried from the GALEX archives (PI program number 079028).  The integration time in the NUV and FUV bands were about 1000 sec.

\section{Results}
\label{sec:results}

\begin{figure*}
 \includegraphics[width=0.9\textwidth]{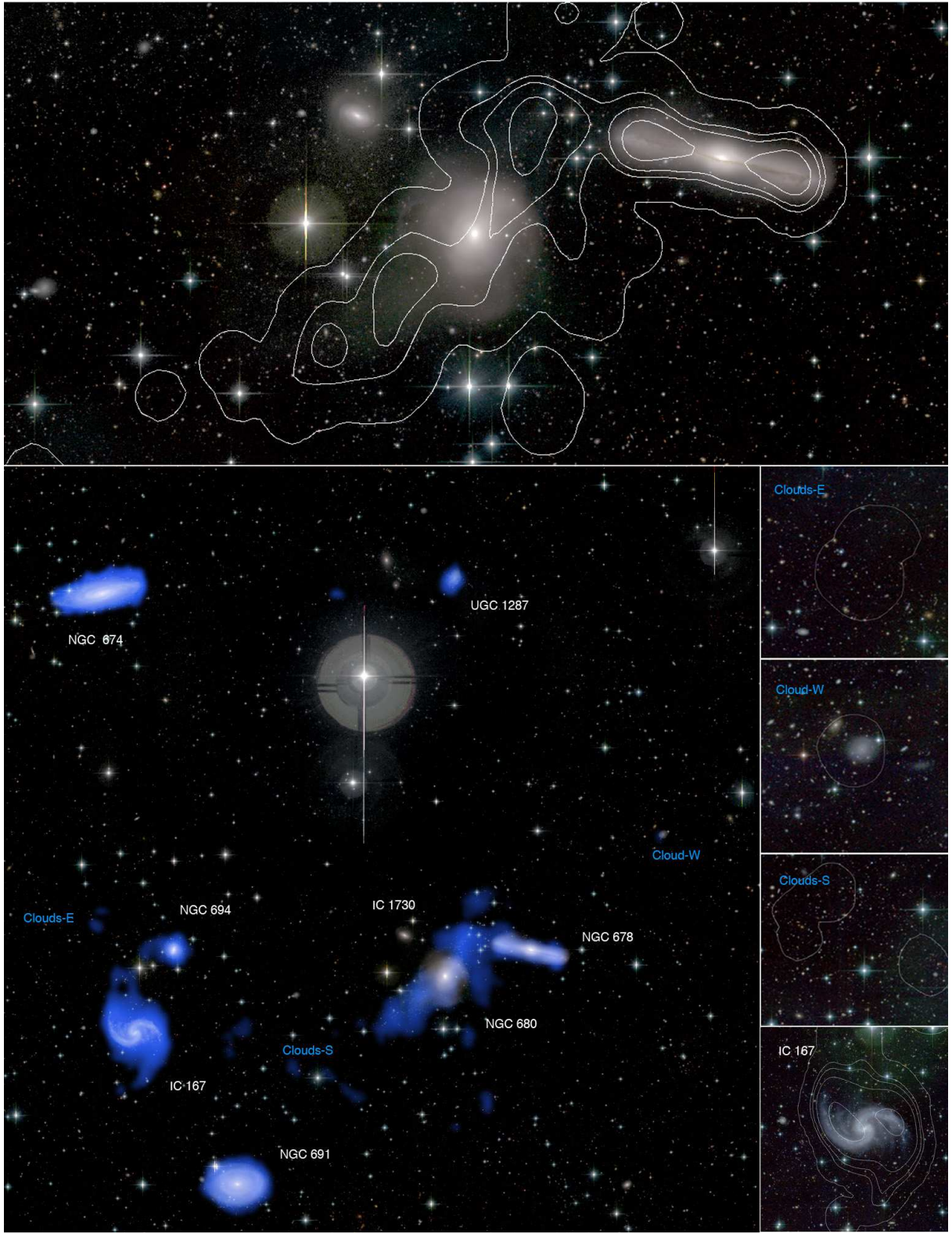}
\caption{{\it Bottom left:} CFHT/MegaCam true-color (composite of the g, r, i bands) image  of the NGC~680 group of galaxies.  The distribution of the HI gas mapped by the WSRT  is overlaid in blue. North is up and East left. The field of view is $49.0'  \x 47.0' $ (535 \x 513 kpc, at the distance of the galaxy). {\it Top:} Close up towards NGC~680 and its closest companions. The contours of the  HI emission (Levels: $\NHI=2,12,20,36\x10^{19}~\cmm$)  are superimposed on the true  color image.  {\it Bottom right:} Close up towards the intergalactic HI clouds in the group and the interacting galaxy IC~167. }
\label{fig:N680-main}
\end{figure*}

\subsection{Detection of new fine structures}
The presence of some level of disturbances had already been reported in the two galaxies under study. 
Looking at  CCD images, \cite{Ebneter88} noted the possible presence of ``patchy dust" in NGC~680 while the structure of NGC~5557 is described as chaotic.  \cite{Schweizer90} derived a rather low but non zero $\Sigma$ index  for the latter galaxy. 
As a matter of fact, hints for asymmetric structures in the outskirts of both ellipticals are barely visible on the shallow INT and SDSS images that were available before the CFHT observations presented here.

The ultra-deep  MegaCam images revealed that the previously detected  disturbances were real, and furthermore disclosed a wealth of so far unknown fine structures, such as shells and long tidal tails. They may be seen on the true color (g,r,i) images   shown in Fig.~\ref{fig:N680-main} and Fig.~\ref{fig:N5557-main}  as well as on the single-band  surface brightness maps  shown in Fig.~\ref{fig:N680-sbmap} and Fig.~\ref{fig:N5557-sbmap}. For future reference, the  fine structures are labelled on the images shown in Fig.~\ref{fig:N680-struct} and Fig.~\ref{fig:N5557-struct}  for which the galaxy models obtained with GALFIT have been subtracted.  

\subsubsection{NGC~680}
\label{sect:N680}

\begin{figure*}
 \includegraphics[width=\textwidth]{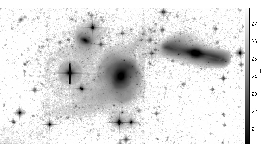}
\caption{CFHT/MegaCam i--band surface brightness map of NGC~680. The surface brightness scale in \sbr\ is shown to the right. The faint foreground stars and distant background galaxies were subtracted from this map.  North is up and East left. The field of view is $17' \x 10' $  (212 \x 125 kpc, at the distance of the galaxy).}
\label{fig:N680-sbmap}
\end{figure*}

\begin{figure*}
 \includegraphics[width=\textwidth]{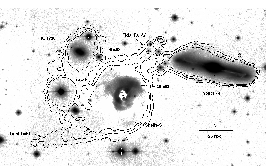}
\caption{g-band surface brightness map of NGC~680, after the subtraction of a galaxy model with a single Sersic index of 4. The faint foreground  stars  and distant background galaxies have been subtracted from the image. The contours of the galactic emission before the galaxy subtraction, as shown on Fig.~\ref{fig:N680-sbmap} , are superimposed  (levels: 28, 27 and 26 \sbr ). The various fine structures disclosed by the image are labeled.}
\label{fig:N680-struct}
\end{figure*}

Extended structures show up on each side  of NGC~680. They are visible in the three observed bands. The halo of a bright star contaminates the emission of the southern stellar plume. Given the color of the star, the effect is however  strongest in the g and r bands. The r-band has the longest exposure time (see Table~\ref{tab:obsmeg}) but suffers from a higher  background level. For this specific system, the i-band    
image shown in  Fig.~\ref{fig:N680-sbmap}  turned out to be the cleanest and most informative one. 
The two stellar plumes to the South-East and North-West of NGC~680 (TT-E and TT-W)  are very broad, diffuse and have medium projected lengths of resp. 75 and 60 kpc. Their surface brightness ranges between 26 and 28.5~\sbr\  in the g--band (25.5 and 28~\sbr in the i band). 
The optical luminosity of the plumes contributes to about 2 percent of the total optical luminosity of the galaxy.

NGC~680 belongs to a group of galaxies and  has several close companions. 
The deep MegaCam images did not reveal any obvious stellar bridge between the elliptical and NGC~678  to the North-West.
However  both galaxies have similar velocities (see Fig.~\ref{fig:N680-field-vel}) and are most likely already interacting. NGC~678 is an edge-on spiral showing a prominent dust lane along its disk, best visible in the  g-i  color map image displayed in Fig.~\ref{fig:N680-g-i}. To the West, the dust lane seems to bifurcate. To the North-East, a  warp is visible on the optical bands.  All these features plus the thickness of the disk  indicate that the spiral is likely disturbed.  Its unfavorable orientation precludes the detection of tidal tails.
The second companion, IC~1730, lies within a common very low-surface brightness envelope, but for this galaxy as well, no clear bridge is distinguishable. \\

As shown on Fig.~\ref{fig:N680-main}, the stellar plumes match well the extended HI structures mapped by the WSRT. 
The true color image  and g-i color map (Fig.~\ref{fig:N680-g-i})  do not reveal any compact  blue star--forming regions along the plumes, despite their gas richness. 
The existence of gaseous tails around NGC~680 was actually known from previous HI observations by  \cite{vanMoorsel88} who moreover disclosed numerous other intergalactic HI clouds further away from the galaxy.

The  presence of apparently free floating clouds is confirmed by the WSRT data. The HI structures labeled as Clouds-S and Clouds-E do not show any optical counterpart at our sensitivity limit of 29~\sbr while a dwarf galaxy is detected at the location of Cloud W (see insets in Fig.~\ref{fig:N680-main}).
The velocity map of Clouds-E and  Clouds-S (Fig~\ref{fig:N680-field-vel}) suggests that the HI structure is a tidal tail emanating from the interacting galaxies IC~167/NGC~694 rather than from our target galaxy, NGC~680. Indeed there is a continuous velocity field from the galaxy pair to the tip of the HI structure.\\

At smaller galactocentric radius, the number of distinct stellar fine structures increases. One of the most prominent is a narrow arc to the East, which has a maximum surface brightness of 25.5~\sbr\  in the g--band (Arc-E on Fig.~\ref{fig:N680-struct}). This arc is much bluer than the surrounding material (see 
Fig.~\ref{fig:N680-g-i}) and probably made of stars of different age or metallicity than the main body.  Subtracting a model of the elliptical from the image, a few faint concentric shells show up (Shells1,2,3 on Fig.~\ref{fig:N680-struct}). More towards the nucleus,   brighter, asymmetric  structures are disclosed. Whether they are real or an artifact of the model used here is unclear. In any case, the images and color map of the central regions of  NGC~680 suggest a rather high level of disturbances, which might be due to some remaining star-formation activity (see discussion in Sect.~\ref{ref:stars} on that matter).

\begin{figure*}
 \includegraphics[width=\textwidth]{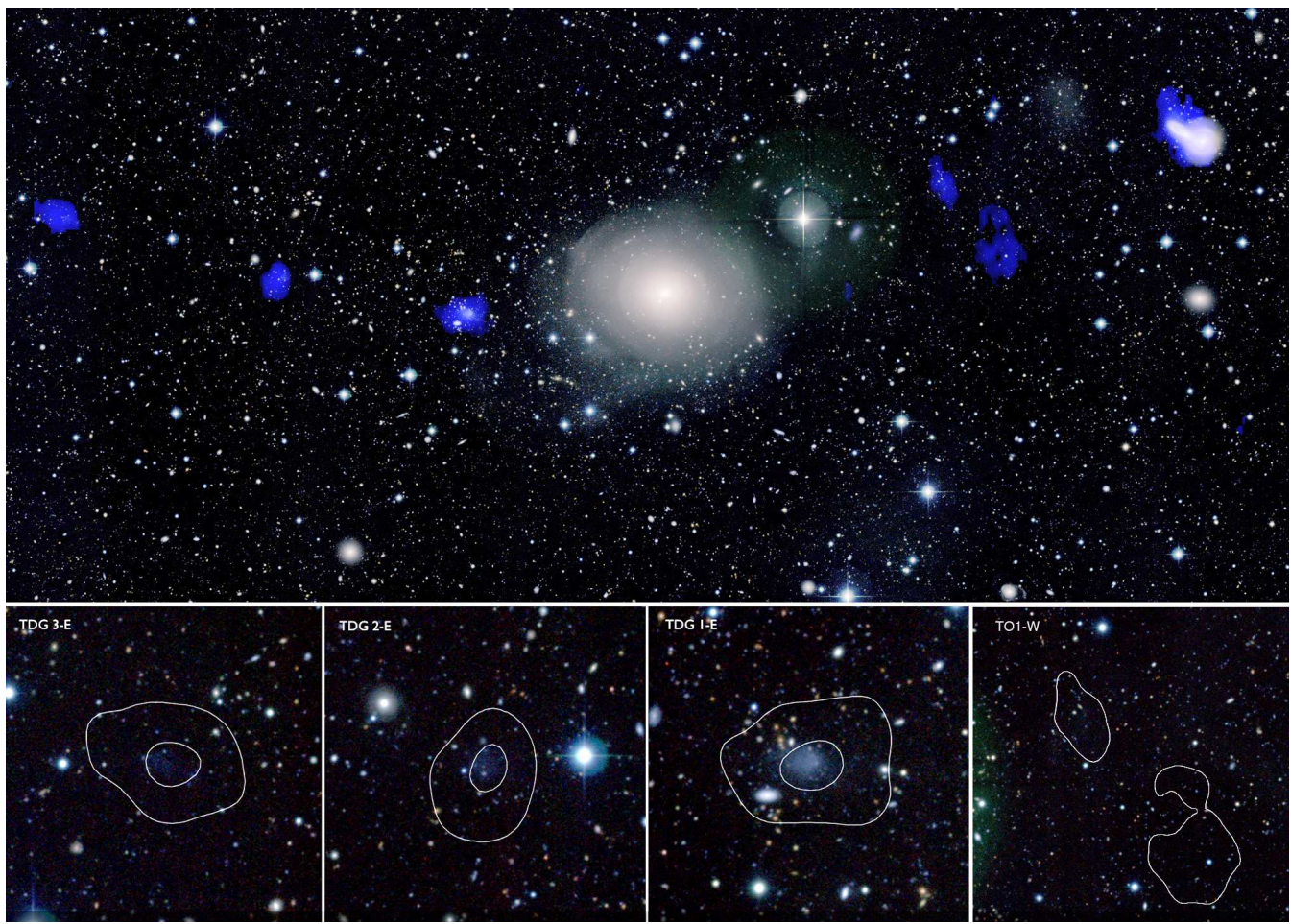}
\caption{{\it Top:} CFHT/MegaCam true-color (composite of the g, r, i bands)  of NGC~5557.  The distribution of the HI gas mapped by the WSRT  is overlaid in blue. North is up and East left. The field of view is $39.5'  \x 18.8' $ (446 \x 212 kpc, at the distance of the galaxy). {\it Bottom:} Close-up towards the intergalactic HI clouds around the galaxy.  Fields of view are 3'\x 3' for TDG1-3-E and 5' \x 5' for TO1-W. The HI contours from the robust weighted map (Levels: $\NHH=3,12\x10^{19}~\cmm$) are superimposed on the true color images (g,r,i). The HI clouds to the East exhibit blue optical and UV  (GALEX) counterparts and are Tidal Dwarf Galaxies candidates. The clouds to the West have extremely faint counterparts (see Fig.~\ref{fig:N5557-sbmap}).  }
\label{fig:N5557-main}
\end{figure*}

\subsubsection{NGC~5557}

\begin{figure*}
\includegraphics[width=\textwidth]{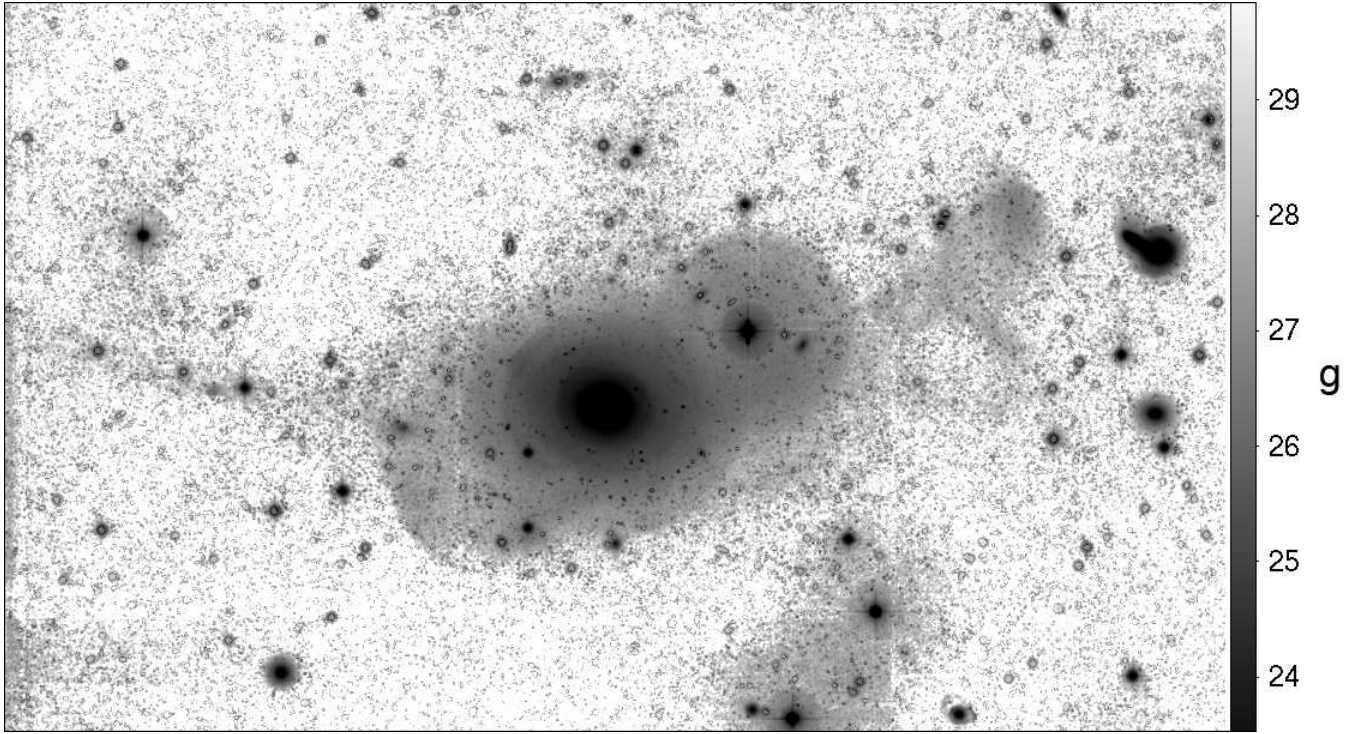}
\caption{CFHT/MegaCam g--band surface brightness map of NGC~5557. 
The surface brightness scale in \sbr\ is shown to the right. It ranges between 23 and 30~\sbr.
The faint foreground stars have been subtracted.
 North is up and East left. The field of view is  $37.0'  \x 22.7' $ (418 \x 256 kpc, at the distance of the galaxy).}
\label{fig:N5557-sbmap}
\end{figure*}

\begin{figure*}
 \includegraphics[width=0.93\textwidth]{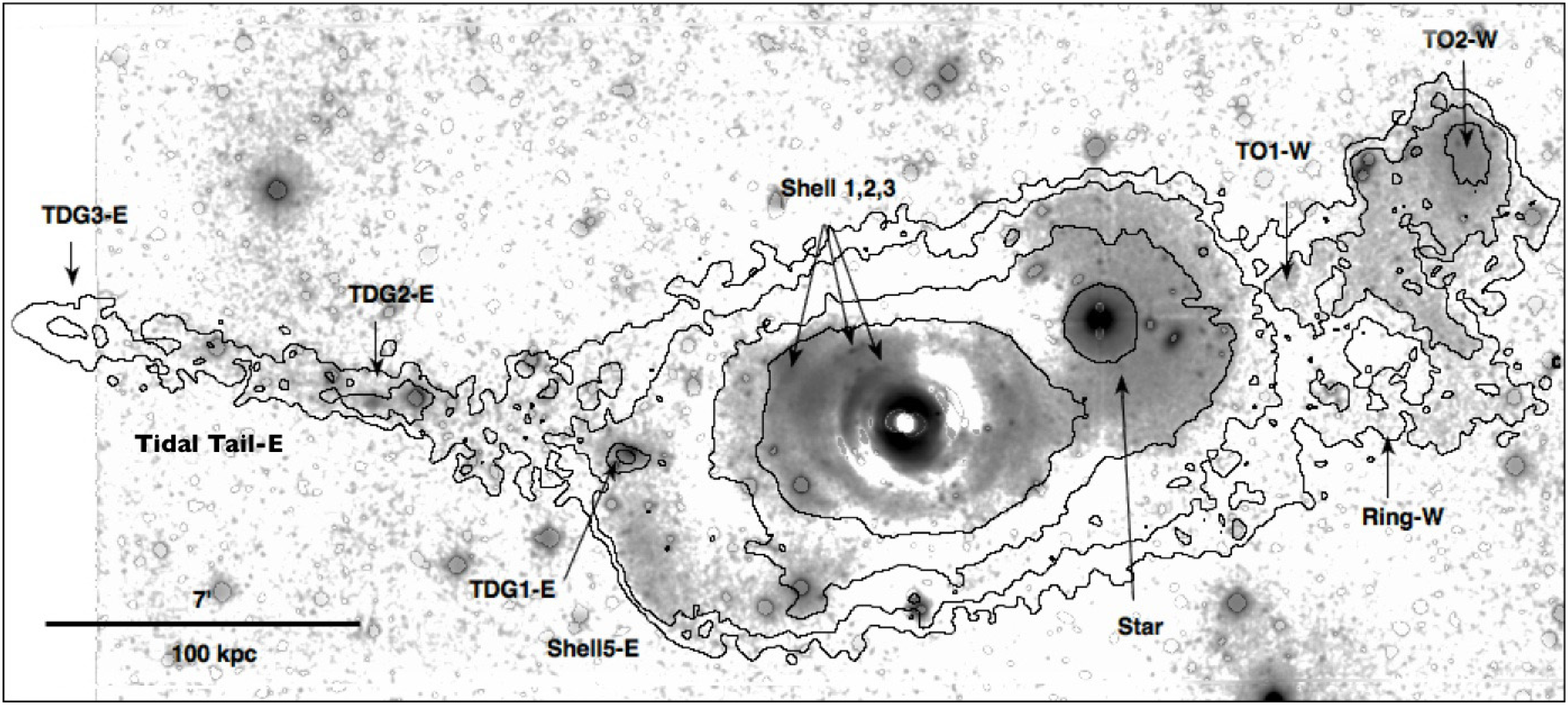}
\caption{g-band surface brightness map of NGC~5557, after the subtraction of a galaxy model with a single Sersic index of 4. The faint foreground  stars  have been subtracted from the image. The contours of the galactic emission before the galaxy subtraction, as shown on Fig.~\ref{fig:N5557-sbmap} , are superimposed (levels: 29, 28, 27 and 26~\sbr). The various fine structures disclosed by the image are labeled.  The image is cut to the East at a location where a high sky background  could not be properly removed (see Fig.~\ref{fig:deep-field}). }
\label{fig:N5557-struct}
\end{figure*}

The most striking optical structures of NGC~5557 are, to the East, a long straight, narrow, filament  (Tidal Tail-E on Fig.~\ref{fig:N5557-struct}), and to the West  a broader, more complex  structure, apparently composed  of an extended object at its tip (TO2-W), a ring (R-W) and a bridge (TO1-W). We will argue in the following that some of these features are the optical counterparts of the apparent free-floating HI gas clouds that had previously been detected around the ETG by the WSRT.

The Eastern tail stretches out from the diffuse halo  around  the elliptical; it is lost 15 arcmin further to the East in a region of the MegaCam frame with a high local background that could not be removed (see Fig.~\ref{fig:deep-field}). Its projected distance is at least 160 kpc.
 The filament is clearly visible in the g (Fig.~\ref{fig:N5557-sbmap}) and r (Fig.~\ref{fig:deep-field})  bands; its average surface brightness is however very low: 28.5 \sbr\ in the g-band. At three positions along the tail, the  
surface brightness increases to 25-26  \sbr; these clumps (TDG1-E,TDG2-E,TDG-3-E) appear to be rather blue on the  true color image shown in Fig~\ref{fig:N5557-main}, and most likely correspond to star--forming regions, as further suggested by their detection in the GALEX far--ultraviolet band.  Each of them is associated to an individual HI cloud (see Fig.~\ref{fig:N5557-main}).  As argued in Sect.~\ref{sec:tdg}, these blue, star--forming, gas--rich, objects host by  a redder  filament are likely Tidal Dwarf Galaxies, which condensed  out of  material expelled during the  galaxy-galaxy collision.

The western structure is much more difficult to interpret. 
Its base towards NGC~5557 is lost within the halo of a bright red star. 
In the g-band image shown in Fig.~\ref{fig:N5557-struct}, there seems to be an extension south of the star, linking the galaxy to object  TO1-W. 
Between the 26  \sbr\  isophote of the elliptical and its western most tip,  the structure has a projected size of 11.9 arcmin, or 130 kpc at  the adopted distance of NGC~5557.   Its surface brightness is at its maximum  towards TO2-W where it reaches  26~\sbr. As shown in  the g-i color map (Fig.~\ref{fig:N5557-g-i}), TO2-W has the same color as the outer region of the elliptical. Besides the presence of a bridge, the similar color suggests that the western structure and the outer regions of the galaxy are composed of similar stellar material, implying  that TO2-W is made of tidal debris as well. Other interpretations are however possible, as discussed in Sect.~\ref{sect:merg}. The ring-like structure R-W is in particular intriguing. West of NGC~5557, TO1-W, and possibly R-W are the only objects detected in HI.

Overall the distance between the two extreme tips of the low-surface brightness structures on each side of NGC~5557 (i.e. between TDG3-E and TO2-W) is 375 kpc, making it, if indeed physically connected, one of the largest stellar structures  ever observed around a galaxy. As comparison, the tidal tails of the well known prototypical merger, the Antennae, have a maximum size of 100~kpc, while the so-called Superantennae, a more distant  merging system which is ultraluminous in the far--infrared,  have a total extent of 350 kpc  \citep{Mirabel91}.\\

  Getting closer to the central regions of the elliptical, several additional features are visible, starting with a shell (Shell5-E), located at 80 kpc from the nucleus. There the surface-brightness suddenly drops below the detection limit of 29~\sbr. Further in, the images shown in Fig.~\ref{fig:N5557-main} and \ref{fig:N5557-sbmap} reveal two shells  (S1,S2) with g-band surface brightness of about 25.5~\sbr. These are probably part of a series of concentric ripples. The inner most ones are disclosed subtracting a model of the elliptical from the image (see Fig.~\ref{fig:N5557-struct}). Artifacts  of the method for the galaxy modeling with GALFIT become prominent in the inner most regions. However shell S3 is also present when subtracting a galaxy modeled by an ellipse fitting algorithm (`ellipse' within IRAF) and is probably real.   

These shells do not show up on the g--i color map shown in Fig.~\ref{fig:N5557-g-i}. There is thus no hint that their stellar population differs from the one of the parent galaxy and comes from an external accreted galaxy containing stars with radically  different age  or metallicity, like for a disrupted dwarf satellite. The bluer shell around NGC~680 is in that respect much different. Despite the presence of faint shells, the central regions of NGC~5557 have overall very regular isophotes, as expected for a fully relaxed galaxy. The study of the most  external regions reveal a much more disturbed history.

\subsection{Characterization of the fine structures}

To summarize, our MegaCam images have allowed to considerably increase the number of detected fine structures both in NGC~680 and  in NGC~5557. To quantify this, we used the  empirical approach of  \cite{Schweizer90}, and computed the fine structure index $\Sigma$, defined as  $\Sigma=S+log(1+n)+J+B+X$, where $S$ is the strength of the most prominent ripples, $n$ , the number of observed ripples, $J$ the number of jets, i.e. tidal structures, $B$ and $X$, two indexes characterizing the inner morphology  (Boxiness  and 'X--structure').  
 \cite{Schweizer90} had already determined it for NGC~5557, based on the shallow CCD images available at that time. The value of 
$S+B+X$ has in principle not changed with the new imaging; $n$ has doubled with the detection of the outer shells, and the number of tidal structures went from 0 to at least 3 (TT-E,TO1/2-W,R-W).  Therefore the value of $\Sigma$ should increase from 2.78, the value determined by  \cite{Schweizer90},  to 6.
NGC~680 was not in the sample of  \cite{Schweizer90}. Nonetheless, we have tried to roughly determine its fine structure index, estimating $S$ to the intermediate value of 2, giving to $n$ a minimum of 3 ``ripples", to $S$ 3 tails and arc, 0 to $B$ (no indication of a boxinesss) and 1 to $X$ (tentative $X$ structure).  With all these  assumptions, $\Sigma$ amounts to 6.6. 
The updated values of the fine structure index bring both galaxies close to that of  the prototypical advanced mergers NGC~3921 (8.8)  and NGC~7252 (10.1) \citep{Schweizer92b}. 
\footnote{Note  that the tidal index is unlikely to change significantly  would the depth of the observations  further increase, as confirmed by numerical simulations  of binary  mergers \citep{Michel-Dansac10}. 
A single major merger produces a limited number of tidal tails, generally 4, 2 for each progenitor. Our deep observations basically allowed to reveal structures that became fainter with time.}

Alternatively, the level of tidal perturbations may be determined using less-subjective parameters  such as the so--called ``Tidal parameter", $T$, defined by  \cite{Tal09} as the residual of the galaxy model subtraction. $T$ quantifies all deviations from regular models obtained for instance by fitting the galaxy with elliptic light profiles. However such a parameter turned out not to be functional for our study as it is not able to disentangle between major and minor mergers. Indeed, the newly discovered very extended tidal tails -- which are clear signposts of a past prominent mass assembly event -- have little impact on the measure of the tidal index, because of their very faint luminosity.

\subsection{Intergalactic HI clouds and Tidal Dwarf Galaxies}
\label{sec:tdg}
The WSRT survey has disclosed several apparently isolated  intergalactic HI clouds around both NGC~680 and NGC~5557. The least massive of them,  East and South of NGC~680 (see Fig.~\ref{fig:N680-main}), with column densities below $10^{20}~\cmm$,   are apparently collisional debris from the on-going interactions in the group, but do not exhibit optical counterparts at our sensitivity. A blue optical object is found towards Cloud-W. Its relative isolation in the NGC~680 group and velocity suggest it is  a pre-existing HI--rich dwarf galaxy.   

On the other hand, the three blue objects aligned along the redder  Eastern tidal tail of NGC~5557 appear as Tidal Dwarf Galaxy candidates. A discussion on their real nature is postpone to Sect.~\ref{sec:tdg-disc}.  
 Their global properties are presented in Table~\ref{tab:tdgs} and their images shown in Fig.~\ref{fig:N5557-main}. The closest to NGC~5557, TDG1-E, lies at the border of  the extended stellar halo of the massive elliptical and the base of its tail. The most distant one, at more than 200 kpc from the galactic nucleus, is located near the tip of the tail. 
The  central surface brightness  of all three objects exceeds 25.3~\sbr; they were therefore barely visible  on previous maps of the region.

Their blue optical color and detection in the far UV (see Table~\ref{tab:tdgs-phot}) tell that the dwarfs exhibit an on--going star--formation activity. The  Star--Formation Rate, as estimated from the  GALEX UV fluxes with the formula SFR(UV) = $1.4 \x 10^{-21}$  L$_{\nu}$ [W Hz$^{-1}$ ] \uSFR \citep{Kennicutt98a}, is modest: 0.005 \uSFR\ for the most active galaxy. In agreement with the presence of young stars, the three dwarfs were detected in HI, with a rather high \MHI/\Lb\ ratio of 2--7  \Mo/\Lo.  

\begin{table*}
\caption{Integrated properties of the TDG candidates in NGC~5557}
\begin{tabular}{lcccccccccc}
Name & RA & DEC & d & V & \sbc & \Mb\ & \MHI & \MHI/\Lb  & SFR(UV) \\ 
& & & kpc & \kms & \sbr\ & mag & $10^8$~\Mo\ & \Mo/\Lo & $10^{-3}$~ \uSFR  \\
(1) & (2) & (3) & (4) & (5) & (6) & (7) & (8) & (9) & (10) \\  \hline 
TDG1-E & 14:18:55.9 &  +36:28:57   & 69.7 & 3252  $\pm$  1 $\pm$  5 & 25.3 $\pm$  0.2  &  -14.05  $\pm$  0.05 & 1.4 $\pm$  0.2 &   2.2  $\pm$  0.4 & 5   $\pm$  0.5 \\
TDG2-E  &14:19:24.4  & +36:30:05   &  134.3 & 3196 $\pm$  1  $\pm$  2 & 26.4 $\pm$  0.3  & -12.6 $\pm$  0.1 & 0.8  $\pm$  0.1 &   5.2 $\pm$  1.1 & 2 $\pm$  0.5 \\
TDG3-E  & 14:19:58.1  &  +36:31:53 &  212.1 & 3165 $\pm$  2  $\pm$  5 & 26.9 $\pm$  0.3  & -12.6 $\pm$  0.1 & 1.2  $\pm$  0.2 &   6.7 $\pm$  1.7  & 4 $\pm$  1 \\ \hline
\multicolumn{10}{l}{\parbox{\textwidth}{{\small (4) Projected distance to the nucleus of the parent galaxy (5) heliocentric velocity, determined fitting the peak of the HI spectrum integrated over the area of the TDG. The first value of the error is estimated  from the Gaussian fit. The second value of the error takes into account the velocity gradient observed along the TDG.  (6) Central surface brightness in the B band (7) Absolute blue magnitude in the B band.  B band values were extrapolated from measurement in the g and r bands using the SDSS calibration  by Lupton (2005). (8) HI mass. The errors given are estimated from the integrated spectra obtained with natural and robust weighted WSRT datacubes.    (9) HI mass to blue luminosity  (10) Star Formation Rate estimated from the GALEX FUV luminosity. }}}\\

  \end{tabular}
  \label{tab:tdgs}
\end{table*}

\begin{table*}
\caption{Photometry  of the TDG candidates in NGC~5557}
\begin{tabular}{lcccccccccc}
Name  & FUV & NUV & g & r & i  \\ 
& \multicolumn{5}{c}{mag} \\ \hline 
TDG1-E &    20.7 $\pm$ 0.1   & 19.8   $\pm$ 0.1  & 18.5   $\pm$  0.05  &  18.0 $\pm$  0.05 &  17.7 $\pm$  0.05     \\
TDG2-E  &   21.6 $\pm$ 0.2   &  20.6  $\pm$ 0.2  &  20.1  $\pm$  0.05 &   19.8 $\pm$  0.05 &  19.3  $\pm$  0.05  \\
TDG3-E  &    20.9 $\pm$ 0.2   &  20.1  $\pm$ 0.2  &  20.1  $\pm$  0.1   &  19.9 $\pm$  0.1     &  19.8  $\pm$  0.1     \\
  \end{tabular}
  \label{tab:tdgs-phot}
\end{table*}

\subsection{Hints from the multi-wavelength data on the central regions}
\label{sect:other}

Our deep MegaCam optical images of the two target galaxies tell about a rich past  accretion/merging history. Could it have been inferred analyzing the complementary multi-wavelength data of the \AD\ survey, available for their central regions? 
In this section we present results based on the survey data taken with the  {\tt  SAURON} integral field spectrograph (Paper I), which provides key information on the stellar populations, age and kinematics. We also use survey observations of the atomic gas obtained at WSRT   \citep{Serra11} and of the molecular gas taken with the IRAM 30m telescope (Paper IV).

\subsubsection{Stellar kinematics}
The dynamical and kinemetric analysis  based on the  \AD\  {\tt SAURON} data  are summarized here: \\

$\bullet$ The stellar velocity map of NGC~680, shown in the inset of Fig.~\ref{fig:N680-vel}, undoubtedly suggests regular  rotation.The galaxy belongs to the \AD\ group ``e" of {\it Disk Rotators} (Paper II).  With a specific angular momentum \LamR\ of 0.46 within half of the effective radius, it is a clear fast rotator (Paper III). 
  There is a misalignment  angle of about 37 degree between the kinematic position angle (359.5 degree) of the stellar component in the inner regions (Paper II)  and  that inferred further out from the HI data. 
  Finally, despite its rather  perturbed central morphology (see previous section), the galaxy does not show any remarkable kinematic feature within the {\tt SAURON} field of view ($33\arcsec \times 41\arcsec$; $6 \times 7.4$~kpc), such as a decoupled core.  This  is however typical of the fast rotators in the \AD\ sample.
 
$\bullet$ In sharp contrast, the stellar velocity map of NGC~5557, presented in the inset of Fig.~\ref{fig:N5557-vel}, only  exhibits weak evidence of rotation. The galaxy  belongs to the \AD\ group ``b" of  {\it Non Disk like Rotators}, with a maximum \Vrot\ of 20.5 \kms. 
With a specific angular momentum at one half effective radius of $\LamR\ = 0.044$  while its apparent ellipticity  is $\epsilon=0.144$, the galaxy matches the criteria to be classified as a slow rotator (Paper III). 
No specific kinematical structure is observed.

\subsubsection{Stellar populations}
\label{ref:stars}
The {\tt SAURON} data allows to determine the principle  characteristics, in particular the age, metallicity and abundance ratio,  of the stellar populations in the central regions of the target ETGs. The method which makes use of the line strength index maps of \Hb, Fe5015 and Mgb and predictions from  single stellar population models  is described in \cite{Kuntschner10} and \cite{McDermid11}. Outside the field of view of {\tt SAURON}, the stellar populations may be constrained using their broad--band color profiles. \\

\begin{figure}
 \includegraphics[width=\columnwidth]{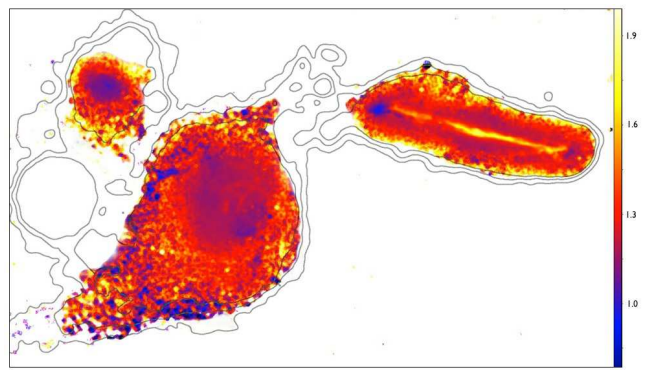}
\caption{g-i color map of NGC~680. All foreground stars have been subtracted. The contours of the surface brightness maps with the same levels as in Fig.~4 are superimposed. }
\label{fig:N680-g-i}
\end{figure}

\begin{figure}
 \includegraphics[width=\columnwidth]{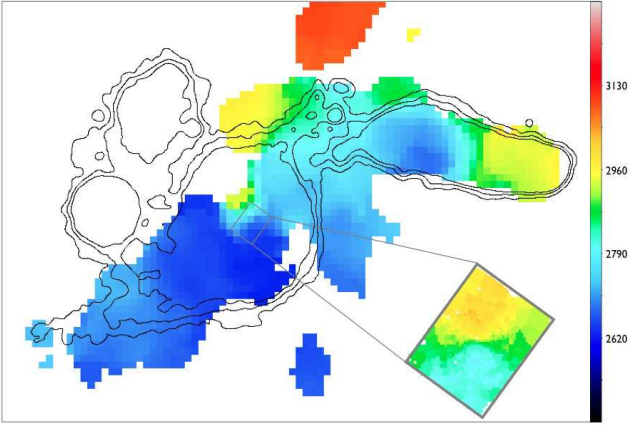}
\caption{Velocity map of NGC~680. The main figure shows the HI velocity map from WSRT while the inset presents the stellar velocity map from {\tt SAURON}. The contours of the surface brightness maps with the same levels as in Fig.~4 are superimposed. The velocity scale in \kms\ is shown to the right. }
\label{fig:N680-vel}
\end{figure}

\begin{figure*}
 \includegraphics[width=\textwidth]{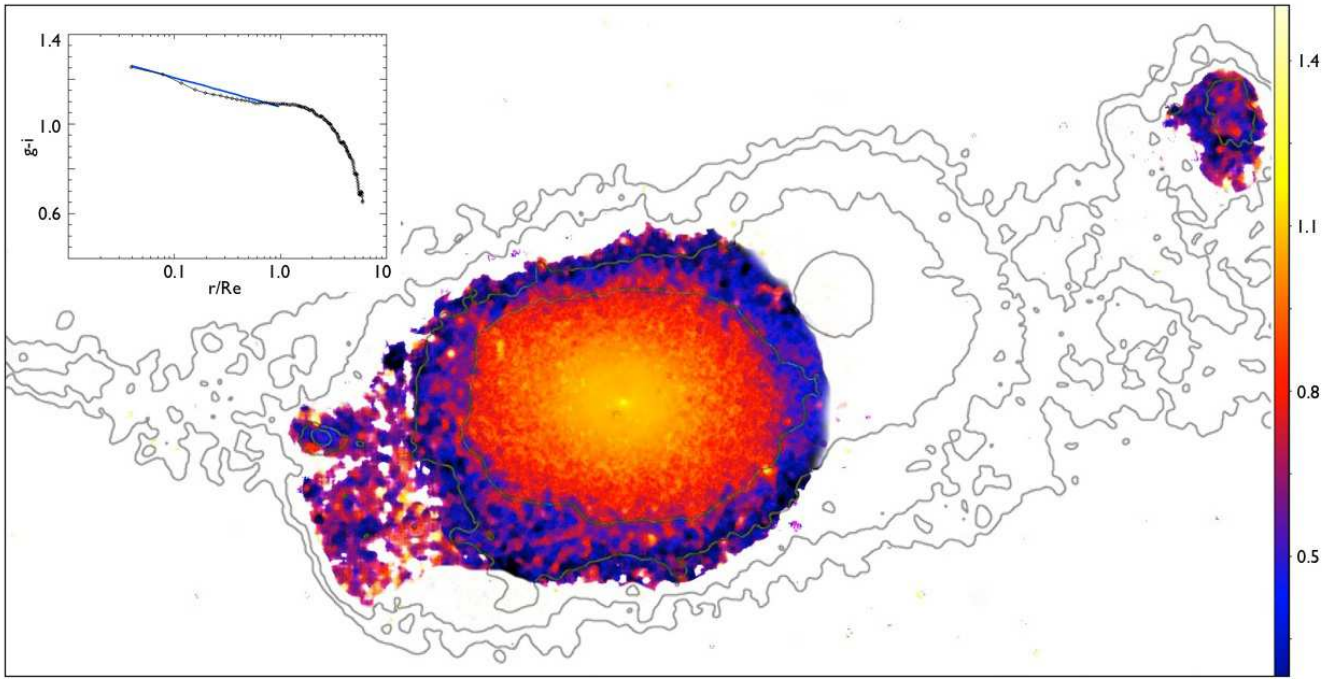}
\caption{g-i color map of NGC~5557. All foreground stars have been subtracted. The contours of the surface brightness maps with the same levels as in Fig.~4 are superimposed. The inset shows the average color profile as a function of radius. The blue curve corresponds to the mean color gradient of a large sample of SDSS ETGs with average $g-i$ colors and $Re$ similar as NGC~5557  \citep{Tortora10}.}
\label{fig:N5557-g-i}
x\end{figure*}

\begin{figure*}
 \includegraphics[width=\textwidth]{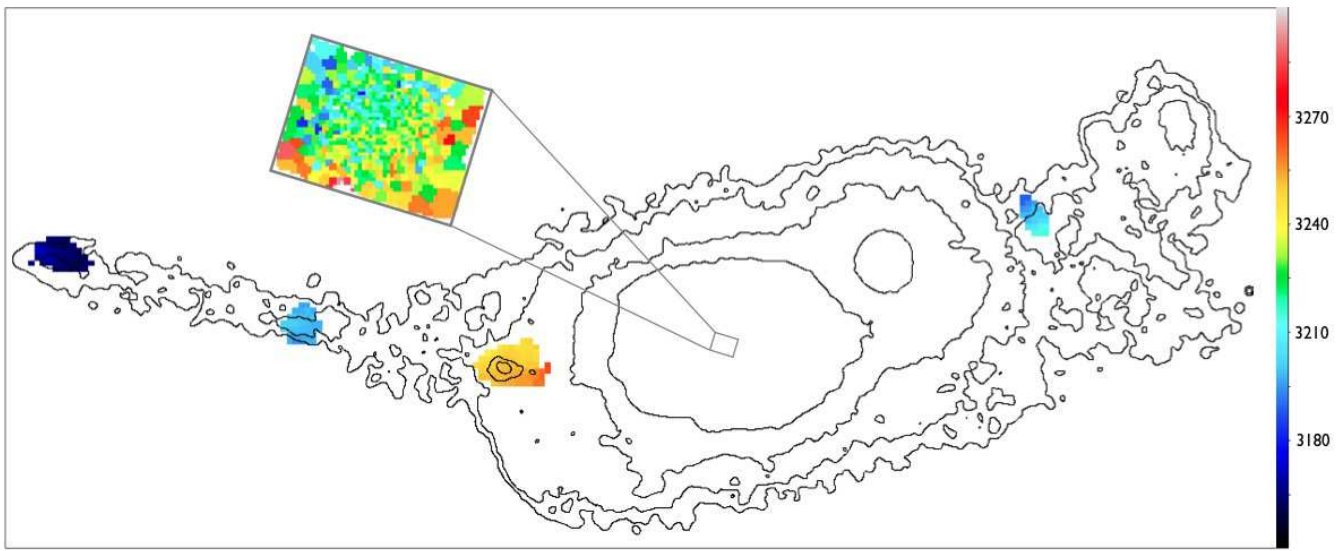}
\caption{Velocity map of NGC~5557. The main figure shows the HI velocity map from WSRT while the inset presents the stellar velocity map from {\tt SAURON}. The contours of the surface brightness maps with the same levels as in Fig.~4 are superimposed.  The velocity scale in \kms\ is shown to the right.}
\label{fig:N5557-vel}
\end{figure*}

\begin{figure}
 \includegraphics[width=\columnwidth]{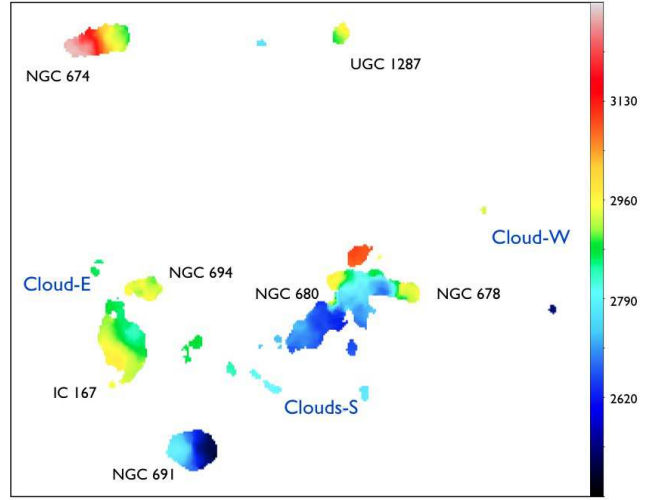}
\caption{Velocity map of the HI emission in the NGC~680 group of galaxies. The intensity scale in \kms\ is shown to the right. The field of view is 0.8 $\times$ 0.7 degree.}
\label{fig:N680-field-vel}
\end{figure}

$\bullet$  The SSP-equivalent mean value of the stellar age of  NGC~680 within the central 0.3 kpc (Re/8) is 7.8 Gyr. It decreases to 6.5 Gyr within the effective radius (14\arcsec\ or 2.6 kpc)  \citep{McDermid11}. The detection by {\tt SAURON}  of ionized gas (as traced by the \OIIIb\ and \Hb\ emission lines) towards the nuclear region might indicate the presence of hot young stars. However,  as recently pointed out by \cite{Sarzi10},   in early-type galaxies most of the ionizing photons are emitted by old stellar populations, in particular the post-asymptotic giant branch (pAGB) stars.
The near and far Ultra-Violet emission detected by GALEX of the core of the galaxy might as well reveal the presence of young stars, but it is now established that one of the main contributors of the UV emission  of  ETGs are old Helium-burning stars \citep{OConnell99}.  Contamination by the UV-upturn \citep[e.g.][ and references therein]{GildePaz07} makes it difficult to determine the SFR of ETGs from their UV emission.
Up to 4 Re,  NGC~680 has a rather uniform red $g-i$ color, with a mild blueing by up to 0.15 mag towards the inner regions (see  Fig.~\ref{fig:N680-g-i}). \\

$\bullet$  The SSP-equivalent   mean value of the stellar age of   NGC~5557 within the central 0.7 kpc (Re/8) is 8.1 Gyr. It decreases to 7.4 Gyr within one effective radius (29\arcsec\ or 5.4 kpc)  \citep{McDermid11}. For comparison, at Re/8, \cite{Denicolo05} estimated an  age of $7\pm1.3$ Gyr, consistent with our determined value.
There is no evidence of the presence of ionized hydrogen towards the nucleus of NG~5557:  the \OIIIb\ and \Hb\ emission lines were not detected by {\tt SAURON}.
 As shown in Fig.~\ref{fig:N5557-g-i}, NGC~5557 exhibits a positive $g-i$ color gradient (i.e. center being redder). Up to  1 Re, the gradient is similar to that measured by \cite{Tortora10}  for a sample of nearby ETGs selected in the SDSS showing the same effective radius and mean color as NGC~5557 ($g-i=1.2$). As argued by  \cite{Tortora10}, the color gradient is likely due to a positive metallicity gradient. Beyond 2 Re and up to 6 Re, the gradient steepens. The average apparent drop by about 0.4 mag is only partly due  to the contamination by the halo of the bright star to the West. According to  Single Stellar Population models \citep{Bruzual03},  if only driven by metallicity, such a gradient would require a drop of metallicity by a factor of at least 100  of the old stellar component. The bluer color in the very outskirts of the galaxy might  reveal  the presence of a population of stars with ages younger than 2 Gyr. Going even further out in the Western halo and collisional debris, the average color increases, except towards the HI--rich star--forming objects located in the tidal tail.

\subsubsection{Gas content}
\label{sect:gas}

The CO(1-0) and CO(2-1) emission lines of the two galaxies were observed with the IRAM~30m antenna as part of the \AD\ survey \citep[][Paper IV]{Young11}. 
They were not detected: the  upper limits for the molecular gas mass are log$(\MHH)< 7.87$~\Mo\  and   log$(\MHH)< 7.92$~\Mo\ for resp. NGC~680 and NGC~5557, assuming a line width of the CO lines of 300~\kms. 
The absence of molecular gas in NGC~5557 is consistent with its low dust content -- the galaxy was not detected at 160~$\micron$ by Spitzer \citep{Temi07} --, and lack of atomic and ionized hydrogen in the central regions. The only HI detected in this system by our WSRT observations  is towards the four star--forming objects along the stellar Eastern and Western  filaments.  Earlier on, \cite{vanDriel01} had obtained an upper limit of log$(\MHI)<9.17$~\Mo\ using the Nan{\c c}ay antenna.

The absence of detected \HH\ in NGC~680 may appear  more puzzling as the galaxy does contain some amount of atomic hydrogen though it is mostly concentrated towards the tidal debris (See Fig~\ref{fig:N680-main}). Within the central beam of IRAM, the HI column density does not exceed $2 \x 10^{20}~\cmm$. 
 $8 \x 10^8 \Mo$ of HI was detected there, giving an upper limit for the mass ratio of \MHH\ over \MHI\  of 1. The central regions of  ETGs usually exhibit much higher \MHH\ over \MHI\  ratios, up to 10 \citep{Oosterloo10}.  \\

We discuss in the following section whether  the properties of the central regions in both galaxies -- absence of KDCs and molecular gas, dominant old stellar populations -- are consistent with the hypothesis of a rather recent mass assembly event,   suggested from the study of the external regions and the detection there of prominent tidal tails.

\section{Discussion}
\label{sec:discussion}

\subsection{A  merger origin for NGC~680 and NGC~5557}
\label{sect:merg}
As summarized  in Sect.~\ref{sect:origin}, the origin of the early--type galaxies has been debated for a long time.
If high-redshift observations of massive elliptical-like galaxies \citep{Cimatti04} seems to be surprisingly well described by a monolithic collapse or a very intense episode of star formation, it is largely agreed  that mergers should have played a key   role in their mass assembly. The issue is in fact to determine when and how  the mass assembly principally occurred: through old or recent binary major mergers, multiple minor mergers or a combination of them.  As clock for age-dating merger events, the fine structures  might bring precious insights to these questions.

The high fine structure index measured in NGC~680 and NGC~5557, close to that of prototypical mergers such as NGC~7252, would suggest a relatively recent major merger for both galaxies. We examine this hypothesis in the following sections.

\subsubsection{Tidal debris vs camera artifacts and foreground objects}
When discussing structures with brightness as low as 28~\sbr, the question about their reality naturally arises. Residuals of CCD gaps in the camera, of background  variations, reflections and ghosts  of bright objects, plus the star halos have indeed similar brightness. However, they usually generate geometric shapes, like grids, disks or very extended diffuse patterns, that differ significantly from that of the fine structures discussed here: narrow tails, shells and arcs, which are typical of collisional debris. Besides, these structures are detected in several bands, which comforts the hypothesis that they are not instrumental  artifacts.

A more  serious issue is the contamination by foreground cirrus. Extended scattered emission from Galactic interstellar clouds and red emission associated with photoluminescence \citep{Witt08} can mimic stellar streams, as recently pointed out by \cite{Cortese10}. Carrying out a multi-wavelength analysis of the colliding galaxies NGC~4435/38 in Virgo, these authors argued that what had previously been taken as a tidal tail in the system was most-likely a Galactic cirrus. 
 The cirrus hypothesis is particularly worth investigating for  the  objects referred  as TO2-W and R-W in Fig.~\ref{fig:N5557-struct}, given their rather puzzling shape.
However,  the IRIS 60/100 $\micron$ maps, queried from the archives of the IRAS satellite,  do not show at these locations the presence of the extended far--infrared emission  expected for a cirrus. The closest FIR source, 4\arcmin\ away, is associated  to the colliding galaxies NGC 5544/45. In the far-UV which is also a reliable tracer of Galactic cirrus \citep{Cortese10}, there is as well no hint for extended UV emission on the 1066 sec exposure FUV image we queried from the GALEX archives. In fact both NGC~680 and  NGC~5557 are  fortunately located in  holes within Galactic interstellar clouds. 
TO2-W and R-W in NGC~5557 are then likely genuine  collisional debris and might correspond to a tidal tail for the former and an arc, perhaps made of stars falling back onto their progenitors for the latter, or tidal structures observed with specific projection angles.

\subsubsection{Minor vs major merger: where does the tidal material come from?}
The tidal structures disclosed by the MegaCam images might either originate from a parent massive galaxy involved in a major merger,  or from  low-mass disrupted companions. 
The huge size of the tidal tails (75~kpc and at least 160~kpc for  resp. the filaments of NGC~680 and NGC~5557) argues for the major merger scenario, i.e. a merger involving progenitors with mass ratios smaller than 1:4. Tidal forces can stretch the stars of  disrupted low mass satellites along large distances  -- e.g. up to 80~kpc  for the stream associated to the Sagittarius dwarf galaxy which is being accreted by the Milky Way  \citep{Martinez04} --, but the tails formed that way usually wrap around the host galaxy, and do not have the linear structure observed for TT-E in NGC~5557. The rare so-called threshing systems, i.e. tidally disrupted in-falling satellites on radial orbits \citep{Forbes03,Sasaki07}, might form long straight tidal tails. Those are however characterized by a remaining red, compact, central core with two tails (a major one and a counter one) emanating from it at different angles. In the case of   TT-E, only one single structure is observed hosting three perfectly aligned blue, diffuse condensations.  
As shown by \cite{Bournaud06}, such Tidal Dwarf Galaxies are unlikely to form in mergers between galaxies with mass ratios above 1:4.  
The broader fuzzy tails on each side of   NGC~680 are as well inconsistent with being the remnants of accreted dwarf satellites.

The Eastern filament of  NGC~5557 resembles the long, star--forming,  tidal tails observed in real and  simulated prograde encounters like in the Antennae \citep{Karl10},  Super-Antennae  \citep{Mirabel91} or the Mice Galaxy \citep{Barnes04}. Seen  edge-on, tidal tails appear as narrow linear structures. 
Finally, the simultaneous presence of two major  tails on each side of NGC~680 and NGC~5557, and more in general their high fine-structure index,  is totally consistent with the hypothesis that they were formed during a merger event including two spiral progenitors.

The nature of the least extended fine structures, such as the sharp-edge arclets and shells observed closer to the main bodies, is much more ambiguous. Phase wrapping  of galaxy companions  produces shells; radial collisions create series of interleaved ripples; the falling-back of tidal material in a major merger can generate shell-like structures as well  \citep[see][for a review]{Struck99}.
The analysis of the stellar populations might help  to  identify the culprit. The arclet East of NGC~680 is much bluer than the surrounding material and might come from stars pulled out from a companion with different ages and metallicities. On the other hand, the shells of  NGC~5557 no longer show up in the color map of the galaxy, implying that they share the same stellar material.

\subsubsection{Old vs recent, wet vs dry, mergers: how to account for the absence of a nuclear starburst?}
Assuming that NGC~680 and NGC~5557 have indeed experienced  a major merger, what constraints do we have on its characteristics and age?

The fact that the tidal tails have not yet vanished gives upper limits for the merger age. The rate at with collisional material evaporates or returns to their progenitors has been quantified by  \cite{Hibbard95b}: 
 according to their numerical simulations of the prototypical merger NGC~7252, more than 80\% in mass of the material located in the tails at the time of the merger will fall back within the following 2.5 Gyr. At time of observations, 0.6 Gyr after the merger, about 3\% of the optical luminosity -- hence of the stellar mass assuming a constant M/L ratio --, emanates from the tidal tails \footnote{The photometry was carried out on ESO/ NTT images of the system (Belles et al.,  in prep).}. For comparison, we estimated from the MegaCam images 
that  the collisional debris  represent about 2--4\% of the stellar mass of  NGC~5557
    \footnote{The measure is rather tricky given the low surface brightness  extension of the tidal tails.   We used  the detection algorithm {\tt MARSIAA} \citep{Vollmer11} which distributes pixels  among different statistical classes, corresponding to the sky, to stars, and  to the LSB objects. Aperture photometry was carried out on the masks generated by  {\tt MARSIAA}. Further details on the method are given in \cite{Ferriere10}. Errors are mainly due to the inaccurate subtraction of the halos of the bright stars and may reach a factor of two in flux.}. 
This relative high value, close to that of NGC~7252, is a strong hint that the merger cannot be too old. To be conservative, an upper limit for  the age of 5 Gyr may be given, corresponding to a merger occurring at a redshift below 0.5. 

Furthermore, a major merger occurring at a redshift well above 0.5 would have left enough time for additional  (minor) merger episodes; they would have contributed to destroy the tails. Analyzing the zoom-in cosmological simulations of  \cite{Martig09}, we have confirmed that even tidal tails formed in major accretion events do not seem to survive  multiple collisions (Michel-Dansac et al., in prep). \\

The detection of gas in the collisional debris indicates that the parent galaxies were gas--rich as well.  NGC~680 contains as much as $3 \x 10^9~\Mo$ of HI,  a large fraction of it being distributed along the tidal tails. In the case of NGC~5557, the HI is now exclusively concentrated in  star--forming objects located in the tidal tail. Their HI gas content (around $10^8~\Mo$) is typical for  Tidal Dwarf Galaxies. The very presence of  TDGs  suggests that their host tail used to be gas--rich.  Indeed TDGs are essentially born from the collapse of gas clouds pulled out from parent merging galaxies, and not from stellar aggregates, as discussed in Sect.~\ref{sec:tdg-disc}.  One may then conclude that the merger responsible for the mass assembly of the galaxies under study was a gas--rich one.  The claim by  \cite{vanDokkum05}  that red  ETGs might be  the result of recent gas--poor mergers has been challenged by  \cite{Sanchez-Blazquez09} and \cite{Serra10}. Our observations raise further doubts about it. 
In any case, the narrow tidal tail observed East of NGC~5557 cannot be generated in the dynamically hot progenitors  that are usually considered for dry mergers. The latter produce fans and plumes rather than tails, as emphasized by \cite{Feldmann08}.\\

The wet merger hypothesis has however some issues. During the merger a large fraction of the atomic gas is channelled onto the central regions where, transformed into molecular gas,  it generally fuels a nuclear starburst though its strength and duration might depend on the geometry of the encounter \citep{DiMatteo08}.  Stellar populations with ages coincident with that of the collision are then expected in the merger remnant. 
However, no molecular gas has been detected in the nuclear regions of the galaxies (\MHH $ < 10^8~\Mo$) and the luminosity-weighted age of their central stars is as old as $\sim$  8 Gyr.
In comparison,  \cite{Dupraz90} measured as much as $3.6 \times 10^9~\Mo$ of \HH\ in the  prototypical merger NGC~7252 and the   age of the star clusters located in its core  was determined by \cite{Whitmore97} to 0.6$\pm0.2$ Gyr, a value in good agreement with the age estimated for the merger with numerical simulations \citep{Hibbard95b}. 

How can these apparently contradictory pieces of information be reconciled? One hypothesis  is that the parent galaxies were not so gassy,  because they  had lost part of their original gas due to the interaction with their environment. Alternatively their initial gaseous disk was very extended, with an HI distribution biased towards the external regions that contribute to enrich the tidal tail but not so much the central regions. 

A more likely hypothesis is that the merger is at least   1--2 Gyr old and that the merger--induced central starburst has used up all the available molecular gas. Gyr old instantaneous  starbursts  would have left little imprint on the stellar populations probed by optical spectrographs. In fact, following the recent work by \cite{Vega10} who studied a sample of ETGs with the mid-IR spectrograph IRS onboard Spitzer, one may predict an observational signature of an ancient starburst: the emission of PAH molecules arising from carbonaceous material released by carbon stars formed at the time of the merger. 
No IRS spectrum exists for our targets, but in a not too distant future, the JWST should be able to check this prediction.

\subsubsection{The internal structural parameters: consistent with predictions for major mergers?}
\label{sec:kin-maj}
If NGC~680 and NCC~5557 seem to share a common origin -  a  major  merger that left prominent fine structures --, their internal kinematics diverge.  One is a fast rotator (NGC~680) and one a slow one (NCC~5557), which in the \AD\ classification scheme makes them fundamentally different systems \citep{Cappellari11}. Are numerical models of major mergers able to produce objects as similar and at the same time as different as these two galaxies? 

According to the analysis of  the morphology and kinematics of about 100 binary merger remnants presented in Paper VI,   major mergers with mass ratio 2:1 and 1:1  can form fast and slow rotators, while minor mergers of spirals only produce fast rotators. 
Fast rotators are naturally produced  when the remnant manages to keep the initial angular momentum of the most massive companion  or when the colliding galaxies have orbits  that generate some angular momentum in the remnant; this is the case for prograde encounters between equal mass spirals.  This condition is relaxed if one of the colliding galaxies is a late-type spiral. This is a likely hypothesis for NGC~680, which exhibits diffuse, plume--like, tails more typical of retrograde encounters, but has at the same time a rather  high gas content, suggesting that at least one of its  progenitor was  gas--rich and thus of a late type.
On the other hand, slow rotators are formed when at least one of the merging galaxies is on a retrograde orbit with respect to the other, to cancel out the  angular momentum  \citep{Jesseit09}. As noted in Paper VI, the two progenitors do not need to be on retrograde orbit; thus the long edge-sharp tidal tail observed East of NGC~5557, which is typical of prograde collisions, is still consistent with the model,  provided that the second parent had a retrograde orbit. As a  matter of fact, the western tidal structure is thicker, which is  expected  for a  retrograde interaction \footnote{Note that the two main tidal tails of merger remnants come each from one of the progenitors.}. \\

Another result of the simulations presented in  Paper VI  is that slow rotators produced by retrograde  major mergers of spirals  should consist of counter-rotating disks (showing  thus a double $\sigma$ kinematical profile, Paper II) with  Kinematically Decoupled Components made of old stars in their centre. Such KDCs should be absent  in  fast rotators. These predictions are in agreement with the observations for NGC~680, but not for NGC~5557 which does not exhibit the expected  kinematical signatures.  
A KDC might however be invisible if very small or observed with an unfavorable projection (see Fig.~2 in Paper VI).  It might furthermore  be destroyed if the merger remnant suffers later on additional  collisions. But as argued earlier, such collisions should as well contribute to wipe out the tidal tails.\\

Reproducing the round shape of NGC~5557  is another challenge. 
Unless the impact parameter is very low  \citep{Hoffman10},  binary major mergers  usually produce remnants  that are rather elongated. To further shape  the round body of the most massive slow rotators, such as those observed in clusters of galaxies,  minor accretion events following the major one might be needed (Paper VIII). Thus the initial merger event should have occurred long ago to give enough time for these final arrangements. Modeling the metallicity and stellar populations of the putative disky progenitors,  \cite{Naab09} also claimed that the merger at the origin of present day ETGs should be at least 3--4 Gyr old. 
 As a last possibility,  the galaxy might be a rare case of nearly face-on fast rotator, with the low rotation being due to inclination. Investigating a statistical significant number of similar cases would be required to assess whether this is a sensible option.

In any case, a complex mass assembly history initiated by a major old merger followed by several minor mergers does not seem consistent with the properties of   NGC~5557, as deduced from the MegaCam observations: the presence of long tidal tails argues for a $z<0.5$ merger and  no clear imprints  of recently disrupted satellites, like arclets and small tails are visible.

 \subsubsection{The  large--scale environment:  favorable to multiple mergers?}

The large--scale environment impacts the mass assembly of galaxies, in particular if multiple mergers play a role.

 As shown in Fig.~\ref{fig:N680-main}, NGC~680 belongs to a rather compact group of galaxies; at least 6 gas--rich companions are located within a distance of 300~kpc from the galaxy. According to the $\Sigma_3$ environment density estimator  \citep[][Paper VII]{Cappellari11b}, which appears to correlate best with galaxy morphology, with  $\Sigma_3=1.47$ Mpc$^{-2}$, NGC~680 has the densest environment explored by our survey.  
 The closest massive  galaxy lies at less than 50~kpc from its nucleus and  shows evidence of a tidal interaction  (see Sect~\ref{sect:N680}). 
  Clearly, NGC~680 has had  many opportunities to accrete gas and stellar material from several companions of various masses, though again a major merger is most likely required to form its two prominent tidal tails. In particular the arc East of the galaxy, with its  color bluer than the surrounding stellar population, might correspond to a disrupted  satellite which was recently accreted.

NGC~5557 has a lower $\Sigma_3=0.30$ Mpc$^{-2}$. It does however also belong to a group. At least 13 group members were identified in the census of \cite{vanDriel01}. Their rather low velocity dispersion (about 50~\kms) favors tidal interactions and mergers. Contrary to NGC~680, NGC~5557 is by far the most massive galaxy in the group (see Fig.~\ref{fig:deep-field}).
The lack of very nearby companions and thus the low probability of further minor merger events might explain why the narrow tidal tail of NGC~5557 has survived until now, if formed in a major merger that occurred a couple of Gyr ago.
The location of NGC 5557 in a moderately  dense environment might however be relevant to explain one of its intriguing properties not well accounted for by the numerical models discussed above: its round shape.  Preliminary studies indicate that the potential well of groups and clusters modify the orbits of the galaxies in such a way that   collisions with a low impact parameter  become more probable than in the field (Martig et al., in prep.). This contributes to enhance the merger induced star--forming activity in interacting  galaxies \citep{Martig08} and at the same time produce rounder merger remnants  \citep{Hoffman10}.

\subsection{The major merger scenario and the formation of Tidal Dwarf Galaxies}
\label{sec:tdg-disc}
The formation in collisional debris of objects with masses similar to that of dwarf galaxies is a by-product of major mergers. Tidal Dwarf Galaxies (TDGs) are usually observed along long tidal tails. They appear as gravitational bound gaseous and stellar systems that are  kinematically decoupled from their parent galaxies. Numerical simulations are able to form them  and give clues on their formation mechanism. Several formation mechanisms have been proposed: local gravitational instabilities in the stellar component  \citep{Barnes92} or gaseous component \citep{Wetzstein07}, ejection of Jeans-instable gas clouds \citep{Elmegreen93}, a top--down scenario in which the large--scale shape  of tidal forces play a major role  \citep{Duc04b}, fully compressive mode of tidal forces \citep{Renaud09}, merger of Super-Star-Clusters \citep{Fellhauer02}.  The long term survival of TDGs  has also been questioned. The numerical importance of TDGs among the dwarf galaxy population will depend on their real life expectancy. 
 The TDG candidates discovered in NGC~5557 provide useful constraints on these issues.

\subsubsection{A tidal origin for the star--forming dwarfs?}
The very location of the three objects along what appears as a long tidal tail from an old merger makes them natural Tidal Dwarf Galaxy candidates. Several arguments against them being pre-existing dwarfs and thus in favor of a tidal origin are summarized here: \\
$\bullet$
 Their velocity, as determined from the  21~cm HI line, is within 50 \kms\ of the systemic velocity of  NGC~5557 (3219 \kms). Such a velocity difference is in  the range  predicted by numerical simulations of the formation of TDGs \citep[see Fig.~10 in ][]{Bournaud06}. Furthermore, the velocity monotonically decreases along the tidal tail, consistent with expectations from streaming motions \citep{Bournaud04}.\\
$\bullet$
The dwarfs which have a uniform blue color are hosted by a red tail  likely made of an older stellar population: the g-r color of the tail is 1.3 mag redder than that of the star--forming TDG candidates. Had the tails been produced by tidally disrupted pre-existing dwarfs, their stellar populations should have been similar in the main body and tail regions. 
With such an hypothesis, 	one should as well explain why all the tidal tails from the three torn out dwarfs  are perfectly aligned.\\
$\bullet$
The SFR and \MHI/\Lb\ ratios of the objects are typical of evolved TDGs like those present in NGC~4694 \citep{Duc07b}.\\
Although there is little doubt about the tidal origin of TDG1--3-E, optical spectroscopy allowing a measure of the oxygen abundance in their HII regions  might give a final answer. Indeed, TDGs are deviant objects in the mass-luminosity relation, being too metal rich for their mass \citep{Weilbacher03}.  The detection of molecular gas, as traced for instance by CO, is likely given the high detection rate of TDGs \citep{Braine01}. The lack of dark matter is finally also expected, but difficult to check with observations \citep{Bournaud06}.

Conversely, if these TDGs are confirmed by the independent methods described above, their discovery would be a further hint that NGC~5557 is the result of a wet major merger.

\subsubsection{Long--lived TDGs: numbers and properties}
As discussed earlier tidal material in mergers unavoidably falls back onto their progenitor; so do the TDGs, however at a time scale which depends on their distance to their parent galaxies and orbital motions around their parent  \citep{Bournaud06}. Being located at distances of at least 70 kpc from their host, the TDGs around NGC~5557 may be long--lived. What is their real age? The tail which hosts them has an estimated age of 2--5 Gyr.  A recent numerical simulation of a binary merger \citep[][Belles et al., in prep.] {Bournaud08} indicates that TDGs are formed very early in collisional debris. Therefore, most probably the dynamical age of the TDGs is at least  2 Gyr old. The vast majority of confirmed TDGs known so far has been identified in young, on--going, mergers. An exception is VCC~2062 in the Virgo Cluster \citep{Duc07b}. The discovery of 3 new (candidate) very old TDGs related here is a further proof that such second-generation objects might be long-lived and contribute to the population of satellite dwarf galaxies. Determining their exact number would be of cosmological significance. 

For lack of observations and resolution in numerical models \citep[see though][]{Recchi07b}, the structural properties of old TDGs are ill constrained. When observed at their formation stage, they are remarkably similar to Blue Compact Dwarf Galaxies  \citep{Duc98b}. Will they then evolve into dwarf Irregulars (dIrrs) if they keep their gas, be transformed into Ultra Compact Dwarf Galaxies (UCDs) -- with whom they share a lack of dark matter and perhaps a similar origin in mergers \citep{Chilingarian10} --, become gas--poor dwarf ellipticals (dEs) or even fade  into dwarf Spheroidals, similar to the satellites of the Milky way, for which a tidal origin has been speculated \citep{Metz07}? Interestingly enough, the three old TDGs near NGC~5557 plus VCC~2062 appear as still gas--rich, blue objects, with however a very low central surface brightness (unlike BCDGs, dIrrs or UCDGs). 

\begin{figure}
 \includegraphics[width=\columnwidth]{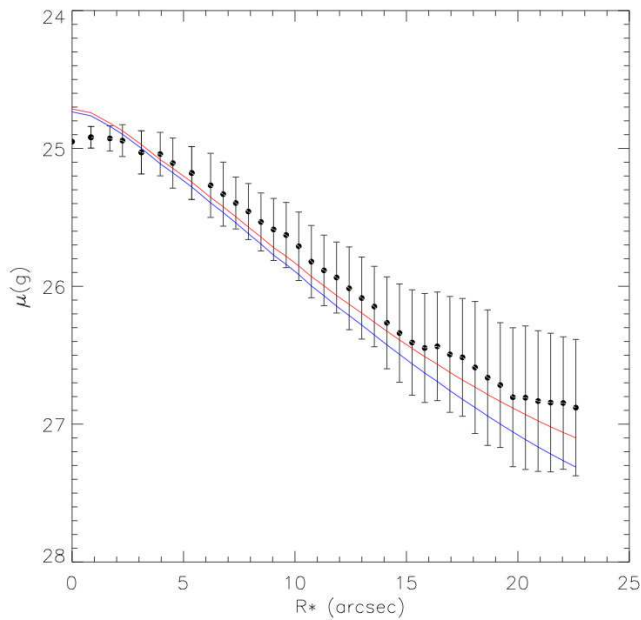}
\caption{Surface brightness profile of TDG1-E, the most luminous Tidal Dwarf Galaxy hosted by NGC~5557. The surface brightness in the g-band (in \sbr) is plotted as a function of the radius in arcsec. The black dots correspond to the observed data; the  blue curve is the model profile from GALFIT with a Sersic index of 0.9; the red curve is the sum of the TDG profile and the profile of the elliptical halo in which the TDG lies.  }
\label{fig:TDG1-E}
\end{figure}

For the most luminous TDG, the observed radial profile (see Fig.~\ref{fig:TDG1-E})  could be fit by GALFIT  with a model characterized by a Sersic index of 0.9 and an effective radius of 11~arcsec (2.1~kpc).  The modeling included two components: the dwarf galaxy itself and the halo of the parent galaxy. 
The  radial profile of TDG1-E  is close to the exponential one  measured in traditional dwarf elliptical galaxies. On the absolute magnitude \Mb\  vs effective radius  \Reff\ diagram, TDG1-E is located within the locus occupied by dEs \citep[see Fig.~1 in ][]{Misgeld11}, though   on the high side of \Reff. 

The deep imaging program carried out as part of the \AD\ and NGVS projects might contribute to identify other old, long-lived TDGs and get more statistically significant information on their properties.

\subsection{Are NGC~680 and NGC~5557 representative of their sub-class?}
\begin{figure}
 \includegraphics[width=\columnwidth]{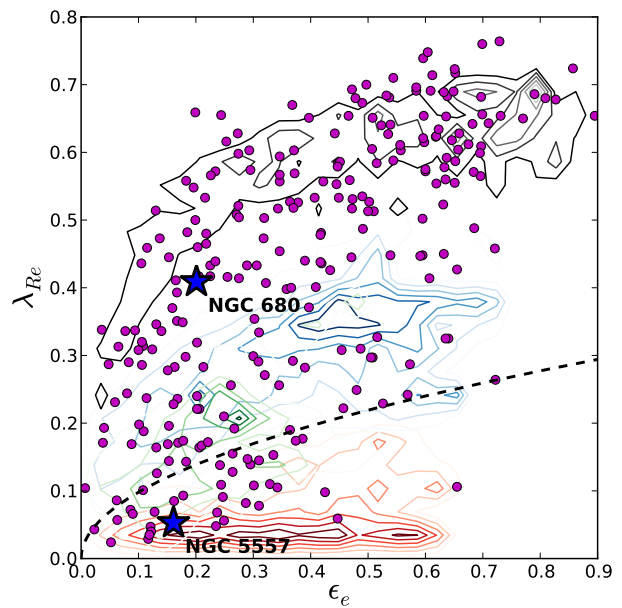}
\caption{Position of NGC~680 and NGC~5557 in the \LamR\ vs $\epsilon$  diagram.  The dots represent the \AD\ ETGs. The contours the distribution of simulated merger remnants for slow (red), fast (blue) rotators, remergers (green) and progenitors (black). Figure adapted from Paper VI.}
\label{fig:Lam-eps}
\end{figure}

NGC~680 and NGC~5557 are only two galaxies among the 260 of the \AD\ sample of nearby ETGs. The question about their representativeness legitimately arises. 
Their location in the reference \LamR\ vs $\epsilon$ diagram is shown on Fig~\ref{fig:Lam-eps}, together with the other ETGs and predictions from the binary merger database presented in Paper VI.

In many aspects, NGC~680 is typical of fast rotators, and more in general of the population of nearby early-type galaxies. Its internal kinematics puts it in the populated class of disk rotators without KDCs (Paper II), to which  82\% of the \AD\  galaxies belong. It contains atomic hydrogen, like about half of the \AD\  ETGs outside Virgo,  and its HI distribution  is  irregular  like half of the HI-detected ETGs \footnote{for the other half, regular HI disks are observed.} \citep{Serra11}
No  molecular gas  was detected in this ETG  like for 76\% of fast rotators (Paper IV).

Massive ellipticals with a round morphology   like NGC~5557 are much rarer among nearby ETGs. Its kinematical group   made of Non Disk-like Rotators with no specific kinematical feature corresponds to only 5~\% of the whole sample. Like the vast majority of slow rotators,  it is not detected in CO, and HI is only found as sparse HI clouds around the galaxy.

Our observations revealed that both objects exhibit prominent gas--rich tidal tails typical of major gas--rich mergers. 
However, as shown in  Fig~\ref{fig:Lam-eps} and discussed earlier, simulations of single binary mergers have difficulties  to reproduce simultaneously their values of  \LamR\  and $\epsilon$. 
We argue here that the binary merger model might be adequate  if very specific initial conditions are considered, such as a low impact parameter for NGC~5557. Such a configuration has statistically a low probability to occur  which would make NGC~5557  itself a very rare system. However as noted in Sect.~\ref{sec:kin-maj} , low impact collisions may be more frequent in dense environments. In that case, NGC~5557  might be representative of slow rotators located in groups and clusters.

\section{Summary and conclusions}

As part of the \AD\ project, ultra-deep MegaCam observations are currently being obtained with the CFHT for  a large sample of about 100 early-type galaxies, complementing a wealth of ancillary data on these systems, including in particular integral field optical spectroscopy, HI/CO mapping and spectroscopy. 
Thanks to a specific  imaging strategy and dedicated data-reduction minimizing the background, detailed here, our images reach a surface brightness limit of 29~\sbr\ in the g--band. Such a sensitivity is unprecedented  for field ETGs.
We have presented in this paper the results obtained for two test cases ellipticals.  Each galaxy belongs to one of the two principle sub-classes of ETGs: the class of fast rotators, which make the bulk of nearby ETGs and the less populated class of slow rotators. Both galaxies had in common the presence of  atomic hydrogen in their close vicinity.

 The MegaCam images revealed  optical counterparts to some previously  detected intergalactic HI clouds and a number of additional fine and diffuse stellar features which were invisible on previously available optical images.
The most spectacular structure is a 160-kpc long filament that sticks out  from the diffuse halo of NGC~5557. It is among the longest known in the nearby Universe.  The tail hosts several blue  star--forming  objects,   detected in HI by the WSRT and in the UV by GALEX. They are likely Tidal Dwarf Galaxies born in collisional debris. 
Two shorter stellar tidal tails with similar morphologies as their HI counterparts are found on each side of NGC~680. 
  More towards the central regions, series of concentric shells and arcs are observed. In NGC~680, a few of these features might be  remnants of  recently disrupted satellites. 
  
Comparing the morphology and kinematics of these structures with predictions from published numerical models of galaxy collisions, we reached the following conclusions:\\

$\bullet$ The tidal tails were formed during a major merger episode. Their shape, internal structure, and high gas content  reveal that the merger  was  a gas--rich one and involved progenitors with mass ratios not exceeding 1:4. 

$\bullet$ The very fact that such tidal tails are still observable, even if very faint,  indicates that all the tidal material has not had the time to fall back or be dispersed by posterior  accretion events and thus that the merger should have occurred at a redshift below 0.5. 

$\bullet$ 
 The striking properties of NGC~5557 -- it simultaneously  exhibits a  high mass, round shape, absence of KDCs, lack of significant rotation, presence of gigantic tidal tails -- do not quite match those obtained with current models of ETG formation.  Cosmological simulations and semi-analytic models  predict that the principle mass assembly  episode  of  massive ETGs, especially the slow  rotators, is more likely to  involve a wet major merger at  a redshift above 1, followed by a long and complex mass accretion history.
 
$\bullet$   In particular, typical binary mergers have difficulties  to directly produce  roundish, massive, slow rotators, unless assuming very specific initial conditions, such as a low impact parameter. We put forward the hypothesis that in rather dense environments like the group which hosts NGC~5557,  collisions with a low impact parameter may indeed be favored. 
 
$\bullet$ Since  the main body appears kinematically  relaxed, the merger cannot however be too recent. This is further confirmed by the  non detection of molecular gas in the core of both galaxies and  absence or limited  star--formation activity there. Whereas NGC~680 and NGC~5557 share many properties with the 0.6 Gyr old prototypical merger NGC~7252, in particular similar fine--structure indexes, they differ with respect to their ability to form stars.
If the merger which formed them triggered a nuclear starburst, as expected for gas--rich mergers, its memory  has already been lost. This pushes the age of the merging episode to at  least 1--2 Gyr, while the conservative upper limit provided by the mass of  the tidal features still observed today may be 5 Gyr.

 $\bullet$ 
  If the merger at the origin of NGC~5557 is indeed a couple of  Gyr old, so are the gas--rich tidal dwarf galaxies discovered along its tidal tail. This would be a proof that they may survive for several Gyr. \\

 This  paper  has shown the great potential of the deep imaging of galaxies  to constrain their origin and  indeed the  in-depth analysis of  NGC~680 and NGC~5557  presented here  seems to challenge the common belief on the mass assembly history of massive ellipticals. Whether these two cases are typical of early-type galaxies, of some sub-classes of them, or are rare unrepresentative objects   will be addressed by our on--going MegaCam survey.

\section*{ACKNOWLEDGEMENTS} 
We thank the referee for very useful suggestions that especially helped improving the readability of the figures. 
We are especially grateful to the CFHT  team for their dedication in the queued service observing and willingness to accommodate specific modes of observation. 
The paper is based on observations obtained with MegaPrime/MegaCam, a joint project of CFHT and CEA/DAPNIA, at the Canada-France-Hawaii Telescope (CFHT), which is operated by the National Research Council (NRC) of Canada, the Institute National des Sciences de l'Univers of the Centre National de la Recherche Scientifique of France, and the University of Hawaii. Images from the GALEX archives and data obtained with the Westerbork Synthesis Radio Telescope, operated by the Netherlands Foundation for Research in Astronomy  ASTRON, with support of NWO, have also been used. 
This research has made use of  the NASA/IPAC Extragalactic Database (NED) which is operated  by the Jet Propulsion Laboratory, California Institute of Technology, under contract with the National Aeronautics and Space Administration.
PAD, FB, EE and LMD acknowledge support from Agence Nationale de la Recherche (ANR-08-BLAN-0274-01). 
MC acknowledges support from a Royal Society University Research Fellowship.
This work was supported by the rolling grants `Astrophysics at Oxford' PP/E001114/1 and ST/H002456/1 and visitors grants PPA/V/S/2002/00553, PP/E001564/1 and ST/H504862/1 from the UK Research Councils. RLD acknowledges travel and computer grants from Christ Church, Oxford and support from the Royal Society in the form of a Wolfson Merit Award 502011.K502/jd. RLD also acknowledges the support of the ESO Visitor Programme which funded a 3 month stay in 2010.
SK acknowledges support from the the Royal Society Joint Projects Grant JP0869822.
RMcD is supported by the Gemini Observatory, which is operated by the Association of Universities for Research in Astronomy, Inc., on behalf of the international Gemini partnership of Argentina, Australia, Brazil, Canada, Chile, the United Kingdom, and the United States of America.
TN, MBois and SK acknowledge support from the DFG Cluster of Excellence `Origin and Structure of the Universe'.
MS acknowledges support from a STFC Advanced Fellowship ST/F009186/1.
NS and TD acknowledge support from an STFC studentship. The authors acknowledge financial support from ESO.

\input{at3dmeg1.rev.v1.bbl}
\end{document}

%% file: abbrev.tex
\newcommand{\MHI}{\mbox{$M_{\mbox{\tiny HI}}$}}
\newcommand{\AD}{\mbox{ATLAS$^{\rm 3D}$}} 
\newcommand{\NHI}{\mbox{$N_{\mbox{\tiny HI}}$}}
\newcommand{\NHH}{\mbox{$N_{\mbox{\tiny H2}}$}}

\newcommand{\Lb}{\mbox{$L_{\mbox{\tiny B}}$}}
\newcommand{\Reff}{\mbox{$R_{\mbox{\tiny eff}}$}}

\newcommand{\LamR}{\mbox{$\lambda_{\mbox{\tiny R}}$}}

\newcommand{\Hb}{\mbox{H${\beta}$}}

\newcommand{\Mb}{\mbox{$M_{\mbox{\tiny B}}$}}

\newcommand{\MHH}{\mbox{$M_{\mbox{\tiny H$_2$}}$}}
\newcommand{\HH}{\mbox{H$_2$}}
\newcommand{\Mo}{\mbox{M$_{\odot}$}}

\newcommand{\Lo}{\mbox{L$_{\odot}$}}

\newcommand{\x}{\mbox{$\times$}}

\newcommand{\Vrot}{\mbox{$V_{\mbox{\tiny rot}}$}}

\newcommand{\Mk}{\mbox{$M_{\mbox{\tiny K}}$}}

\newcommand{\cmm}{\mbox{cm$^{-2}$}}

\newcommand{\kms}{\mbox{km~s$^{-1}$}}

\newcommand{\sbc}{\mbox{$\mu _{\mbox{\tiny B}0}$}}

\newcommand{\uSFR}{\mbox{M$_{\odot}$~yr$^{-1}$}}

\newcommand{\OIIIb}{\mbox{[OIII]$_{\lambda 5007}$}}

\def\sbr{${\rm mag\,\,arcsec^{-2 }}$\ }
\def\sbrs{${\rm mag\,\,arcsec^{-2 }}$}

\newcommand\btab[5]{\begin{table*}[#1]{\parbox{#4}{\caption{#2}}\rule[-0.5ex]{0cm}{0.5ex} }
\begin{tabular*}{#4}{#5} }
\newcommand\sbtab[5]{\begin{table}[#1]{\parbox{#4}{\caption{#2}}\rule[-0.5ex]{0cm}{0.5ex} }
\begin{footnotesize}
\begin{tabular*}{#4}{#5} }
\newcommand{\etab}[4]{
\end{tabular*}
\vspace*{#1}
\begin{flushleft}
\parbox{#2}{#3}
\end{flushleft}
\label{#4}
\end{table*} }

\def\rf@jnl#1{{#1}}
\def\aj{\rf@jnl{AJ }}                   
\def\araa{\rf@jnl{ARA\&A }}             
\def\apj{\rf@jnl{ApJ }}                 
\def\apjl{\rf@jnl{ApJ }}                
\def\apjs{\rf@jnl{ApJS }}               
\def\ao{\rf@jnl{Appl.~Opt.}}           
\def\apss{\rf@jnl{Ap\&SS }}             
\def\aap{\rf@jnl{A\&A }}                
\def\aapr{\rf@jnl{A\&A~Rev.}}          
\def\aaps{\rf@jnl{A\&AS }}              
\def\azh{\rf@jnl{AZh }}                 
\def\baas{\rf@jnl{BAAS }}               
\def\jrasc{\rf@jnl{JRASC }}             
\def\memras{\rf@jnl{MmRAS }}            
\def\mnras{\rf@jnl{MNRAS }}             
\def\pra{\rf@jnl{Phys.~Rev.~A}}        
\def\prb{\rf@jnl{Phys.~Rev.~B}}        
\def\prc{\rf@jnl{Phys.~Rev.~C}}        
\def\prd{\rf@jnl{Phys.~Rev.~D}}        
\def\pre{\rf@jnl{Phys.~Rev.~E}}        
\def\prl{\rf@jnl{Phys.~Rev.~Lett.}}    
\def\pasp{\rf@jnl{PASP }}               
\def\pasj{\rf@jnl{PASJ }}               
\def\qjras{\rf@jnl{QJRAS }}             
\def\skytel{\rf@jnl{S\&T }}             
\def\solphys{\rf@jnl{Sol.~Phys.}}      
\def\sovast{\rf@jnl{Soviet~Ast.}}      
\def\ssr{\rf@jnl{Space~Sci.~Rev.}}     
\def\zap{\rf@jnl{ZAp }}                 
\def\nat{\rf@jnl{Nature }}              
\def\iaucirc{\rf@jnl{IAU~Circ.}}       
\def\aplett{\rf@jnl{Astrophys.~Lett.}} 
\def\apspr{\rf@jnl{Astrophys.~Space~Phys.~Res.}}
\def\bain{\rf@jnl{Bull.~Astron.~Inst.~Netherlands}} 
\def\fcp{\rf@jnl{Fund.~Cosmic~Phys.}}  
\def\gca{\rf@jnl{Geochim.~Cosmochim.~Acta}}   
\def\grl{\rf@jnl{Geophys.~Res.~Lett.}} 
\def\jcp{\rf@jnl{J.~Chem.~Phys.}}      
\def\jgr{\rf@jnl{J.~Geophys.~Res.}}    
\def\jqsrt{\rf@jnl{J.~Quant.~Spec.~Radiat.~Transf.}}
\def\memsai{\rf@jnl{Mem.~Soc.~Astron.~Italiana}}
\def\nphysa{\rf@jnl{Nucl.~Phys.~A}}   
\def\physrep{\rf@jnl{Phys.~Rep.}}   
\def\physscr{\rf@jnl{Phys.~Scr}}   
\def\planss{\rf@jnl{Planet.~Space~Sci.}}   
\def\procspie{\rf@jnl{Proc.~SPIE}}   